\documentclass[journal]{IEEEtran}
\usepackage{xcolor,soul,framed} 
\colorlet{shadecolor}{yellow}
\usepackage[pdftex]{graphicx}
\graphicspath{{../pdf/}{../jpeg/}}
\DeclareGraphicsExtensions{.pdf,.jpeg,.png}
\usepackage[cmex10]{amsmath}
\usepackage{array}
\usepackage{algorithmic}
\usepackage{algorithm}
\usepackage{amsfonts}
\usepackage{amsmath}
\usepackage{bm}
\usepackage{bbm, dsfont}
\usepackage{booktabs}
\usepackage{blindtext}
\usepackage[colorinlistoftodos]{todonotes}
\usepackage{xcolor}
\usepackage{comment}
\usepackage{cite}
\usepackage{diagbox}
\usepackage{enumerate}
\usepackage{enumitem}
\usepackage{eqparbox}
\usepackage[T1]{fontenc}
\usepackage{footnote}
\usepackage{graphicx}
\usepackage{makecell}
\usepackage{mathtools}
\usepackage{mhchem}
\usepackage{mdwmath}
\usepackage{mdwtab}
\usepackage{multirow}
\usepackage{nomencl}
\usepackage{stfloats}
\usepackage{textcase}
\usepackage{threeparttable}
\usepackage{url}
\usepackage{setspace}
\usepackage{subfigure}

\usepackage{etoolbox}
\makeatletter
\patchcmd{\@makecaption}
  {\addvspace{0.5\baselineskip}\egroup}
  {\addvspace{-1\baselineskip}\egroup}
  {}
  {}
\makeatother

\begin{document}

\title{Towards Microgrid Resilience Enhancement via Mobile Power Sources and Repair Crews: A Multi-Agent Reinforcement Learning Approach}

\author{Yi Wang,~\IEEEmembership{Member,~IEEE,}
        Dawei Qiu,~\IEEEmembership{Member,~IEEE,}
        Fei Teng,~\IEEEmembership{Senior Member,~IEEE,}
        and Goran Strbac,~\IEEEmembership{Member,~IEEE}
        
\thanks{This work was supported by two UK EPSRC projects: `Integrated Development of Low-Carbon Energy Systems (IDLES): A Whole-System Paradigm for Creating a National Strategy' (project code: EP/R045518/ 1) and `Technology Transformation to Support Flexible and Resilient Local Energy Systems' (project code: EP/T021780/1), and one Horizon Europe project: `Reliability, Resilience and Defense technology for the griD' (Grant agreement ID: 101075714). \textit{(Corresponding author: Dawei Qiu.)}}
        
\thanks{Yi Wang, Dawei Qiu, Fei Teng, and Goran Strbac are with the Department of Electrical and Electronic Engineering, Imperial College London, London, SW7 2AZ, U.K. (e-mail: \{yi.wang18, d.qiu15, f.teng, g.strbac\}@imperial.ac.uk).}
}

\markboth{IEEE Trans. Power Syst., Accepted for Publication}%
{Shell \MakeLowercase{\textit{et al.}}: Bare Demo of IEEEtran.cls for IEEE Journals}
\maketitle

\begin{abstract}
Mobile power sources (MPSs) have been gradually deployed in microgrids as critical resources to coordinate with repair crews (RCs) towards resilience enhancement owing to their flexibility and mobility in handling the complex coupled power-transport systems. However, previous work solves the coordinated dispatch problem of MPSs and RCs in a centralized manner with the assumption that the communication network is still fully functioning after the event. However, there is growing evidence that certain extreme events will damage or degrade communication infrastructure, which makes centralized decision making impractical. To fill this gap, this paper formulates the resilience-driven dispatch problem of MPSs and RCs in a decentralized framework. To solve this problem, a hierarchical multi-agent reinforcement learning method featuring a two-level framework is proposed, where the high-level action is used to switch decision-making between power and transport networks, and the low-level action constructed via a hybrid policy is used to compute continuous scheduling and discrete routing decisions in power and transport networks, respectively. The proposed method also uses an embedded function encapsulating system dynamics to enhance learning stability and scalability. Case studies based on IEEE 33-bus and 69-bus power networks are conducted to validate the effectiveness of the proposed method in load restoration.
\end{abstract}


\renewcommand{\nomgroup}[1]{%
\ifthenelse{\equal{#1}{A}}{\item[\emph{A.~Indices~and~Sets}]}{%
\ifthenelse{\equal{#1}{B}}{\item[\emph{B.~Parameters}]}{%
\ifthenelse{\equal{#1}{C}}{\item[\emph{C.~Variables}]}{{}}}}
}
\makenomenclature
\setlength{\nomlabelwidth}{1.48cm} 
\nomenclature[A01]{$t \in T$}{Index and set of time steps (hours)}
\nomenclature[A02]{$b \in B$}{Index and set of electric buses}
\nomenclature[A03]{$l \in L$}{Index and set of power lines}
\nomenclature[A04]{$d \in D$}{Index and set of loads}
\nomenclature[A05]{$g \in DG$}{Index and set of diesel generators (DGs)}
\nomenclature[A05]{$g \in PV$}{Index and set of photovoltaic generation (PVs)}
\nomenclature[A06]{$g \in I_{eg}$}{Index and set of mobile emergency generators (MEGs)}
\nomenclature[A07]{$k \in I_{es}$}{Index and set of mobile energy storage systems (MESSs)}
\nomenclature[A08]{$j \in I_{rc}$}{Index and set of repair crews (RCs)}
\nomenclature[A09]{$n \in N_{ms}$}{Index and set of MESS stations (MSs)}
\nomenclature[A10]{$w \in N_{rc}$}{Index and set of damaged components}

\nomenclature[B01]{$\Delta t$}{Time resolution (1 hour)}
\nomenclature[B02]{$c^{ls}_{d}$}{Load shedding cost of load $d$ ({\pounds}/kWh)}
\nomenclature[B06]{$\overline{P}^{eg}_{g}$}{Maximum active power of MEG $g$ (kW)}
\nomenclature[B07]{$\underline{P}^{eg}_{g}$}{Minimum active power of MEG $g$ (kW)}
\nomenclature[B08]{$\overline{Q}^{eg}_{g}$}{Maximum reactive power of MEG $g$ (kVAR)}
\nomenclature[B09]{$\underline{Q}^{eg}_{g}$}{Minimum reactive power of MEG $g$ (kVAR)}
\nomenclature[B10]{$\overline{P}_{k}$}{Power capacity of MESS $k$ (kW)}
\nomenclature[B11]{$S_{k}^{max}$}{Maximum SoC of MESS $k$ (\%)}
\nomenclature[B12]{$\eta_{k}^{esc}$}{Charging efficiency of MESS $k$ (\%)}
\nomenclature[B13]{$\eta_{k}^{esd}$}{Discharging efficiency of MESS $k$ (\%)}
\nomenclature[B14]{$RT^{rc}_{w}$}{Time period required to repair component $w$}
\nomenclature[B15]{$RS^{rc}_{j}$}{Resource capacity of RC $j$}
\nomenclature[B16]{$rs^{rc}_{w}$}{Resources required to repair component $w$}
\nomenclature[B17]{$\overline{P}^{dg}_{g}$}{Maximum active power of DG $g$ (kW)}
\nomenclature[B18]{$\underline{P}^{dg}_{g}$}{Minimum active power of DG $g$ (kW)}
\nomenclature[B19]{$\overline{Q}^{dg}_{g}$}{Maximum reactive power of DG $g$ (kVAR)}
\nomenclature[B20]{$\underline{Q}^{dg}_{g}$}{Minimum reactive power of DG $g$ (kVAR)}
\nomenclature[B21]{$\tilde{P}^{pv}_{g,t}$}{Active power capacity of PV $g$ at time $t$ (kW)}
\nomenclature[B22]{$\overline{S}^{pv}_{g}$}{Apparent power capacity of PV $g$ at time $t$ (kVA)}
\nomenclature[B23]{$\overline{P}^{ed}_{d,t}$}{Baseline of active load $d$ at time $t$ (kW)}
\nomenclature[B24]{$\overline{Q}^{ed}_{d,t}$}{Baseline of reactive load $d$ at time $t$ (kVAR)}
\nomenclature[B25]{$\overline{V}$}{Maximum permissible voltage (p.u.)}
\nomenclature[B26]{$\underline{V}$}{Minimum permissible voltage (p.u.)}
\nomenclature[B27]{$r_{bp}$}{Resistance of line $(b,p)$ (p.u.)}
\nomenclature[B28]{$x_{bp}$}{Reactance of line $(b,p)$ (p.u.)}
\nomenclature[B29]{$\overline{S}_{bp}$}{Capacity limit of line $(b,p)$ (kVA)}

\nomenclature[C01]{$P^{eg}_{g,n,t}$}{Active power generation of MEG $g$ in MS $n$ at time $t$ (kW)}
\nomenclature[C02]{$Q^{eg}_{g,n,t}$}{Reactive power generation of MEG $g$ in MS $n$ at time $t$ (kVAR)}
\nomenclature[C03]{$P^{esc}_{k,n,t}$}{Charging power of MESS $k$ in MS $n$ at time $t$ (kW)}
\nomenclature[C04]{$P^{esd}_{k,n,t}$}{Discharging power of MESS $k$ in MS $n$ at time $t$ (kW)}
\nomenclature[C05]{$S^{es}_{k,t}$}{State of Charge (SoC) of MESS $k$ at time $t$}
\nomenclature[C06]{$u^{es}_{k,t}$}{Binary indicating the scheduling status of MESS $k$ at time $t$ (1 if charging, 0 if discharging)}
\nomenclature[C07]{$Re^{rc}_{j,w,t}$}{Binary indicating if RC $j$ is repairing component $w$ at time $t$ (1 if repairing, 0 otherwise)}
\nomenclature[C08]{$z^{rc}_{j,w,t}$}{Binary indicating if the component $w$ has been repaired by RC $j$ at time $t$ (1 if repaired, 0 otherwise)}
\nomenclature[C09]{$u_{i,n,t}$}{Binary indicating the connection status of mobile unit $i$ on node $n$ at time $t$ (1 if connected, 0 otherwise)}
\nomenclature[C10]{$P^{dg}_{g,t}$}{Active power generation of DG $g$ at time $t$ (kW)}
\nomenclature[C11]{$Q^{dg}_{g,t}$}{Reactive power generation of DG $g$ at time $t$ (kVAR)}
\nomenclature[C12]{$P_{d,t}^{ed}$}{Restored active load $d$ at time $t$ (kW)}
\nomenclature[C13]{$Q_{d,t}^{ed}$}{Restored reactive load $d$ at time $t$ (kVAR)}
\nomenclature[C14]{$P_{g,t}^{pv}$}{Active power of PV $g$ at time $t$ (kW)}
\nomenclature[C15]{$Q_{g,t}^{pv}$}{Reactive power of PV $g$ at time $t$ (kVAR)}
\nomenclature[C16]{$V_{b,t}$}{Voltage magnitude at bus $b$ at time $t$ (p.u.)}
\nomenclature[C17]{$P_{bp,t}$}{Active power flow of line $(b,p)$ at time $t$ (kW)}
\nomenclature[C18]{$Q_{bp,t}$}{Reactive power flow of line $(b,p)$ at time $t$ (kVAR)}
\nomenclature[C19]{$P^{pv}_{g,t}$}{Active power output of PV $g$ at time $t$ (kW)}
\nomenclature[C20]{$e_{b,t}$}{Binary indicating the energized status of bus $b$ at time $t$ (1 if energized, 0 otherwise)}
\nomenclature[C21]{$y_{bp,t}$}{Binary indicating the energized status of line $(b,p)$ at time $t$ (1 if energized, 0 otherwise)}
\nomenclature[C22]{$F_{bp,t}$}{Virtual power flow through branch $(b,p)$ at time $t$}
\nomenclature[C23]{$F^{s}_{a,t}$}{Virtual output of black-start resource $a$ at time $t$}
\printnomenclature

\section{Introduction}
\label{sec:I}
\subsection{Background}
\label{sec:I.A}
Extreme events (e.g., hurricanes, storms, and earthquakes) featured by high impact and low probability can cause a catastrophic effect on power systems (e.g., severe power outages)~\cite{bie2017battling}. Given the serious disruptions, the main goal of a resilient power system should be to maintain the supply continuity of essential loads (e.g., medical facilities and police stations), constituting a system load restoration problem \cite{ding2020power}. As one emerging type of \textit{distributed energy resources} (DERs), \textit{mobile power sources} (MPSs) have been applied in power systems to coordinate with traditional mobile resource \textit{repair crews} (RCs) for system load restoration due to their mobility and flexibility characteristics \cite{yang2020seismic}. However, the deployment of these MPSs and RCs also creates new complications for the resilient operation of power systems, as they are transiting into a decentralized fashion characterized by quick responses as well as various system dynamics and uncertainties \cite{wang2020microgrids}. In this context, it is necessary to develop an efficient decentralized control scheme for the resilient coordination of MPSs and RCs in a complex power-transport network.

\subsection{Literature Review}
\label{sec:I.B}
In the existing literature, the dispatches of MPSs and RCs are commonly solved as individual problems in a centralized manner. On one hand, in \cite{arif2018optimizing,9330590}, RC routing behaviors are formulated as a deterministic optimization problem to address the load restoration in the initial stage, while uncertainties associated with repair time and demand are realized in the later stages. On the other hand, mobile emergency generators (MEGs) have also demonstrated their advanced mobility in restoring essential loads \cite{7559799,9121323} via stochastic programming (SP) considering uncertain line outages. Additionally, mobile energy storage systems (MESSs) as the key technology of MPSs normally coordinate with MEGs for load restoration \cite{wang2022resilience}, while uncertain contingencies are handled via robust optimization (RO) \cite{lei2018routing} and SP \cite{yang2020seismic}. However, prompt load restoration can be influenced by both the operability of a to-be-repaired branch and the availability of MPSs, while effective coordination between RCs and MPSs can significantly enhance the speed and quality of service restoration \cite{lei2019resilient}. It is difficult to achieve the optimal load restoration solely based on RCs or MPSs, leading to the requirement for integrated methods that can coordinate available flexibility resources. To address this issue, there have been a few studies \cite{lei2019resilient,ye2020resilient,ding2020multiperiod} investigating the coordinated effect between MPSs and RCs for system load restoration in coupled power-transport networks. However, the above studies are all deterministic given the difficulty of modeling the complex stochasticity of MPS and RC operations in a centralized manner. In \cite{zhang2021stochastic}, a two-stage stochastic pre-allocation model is proposed to coordinate MPSs and RCs, capturing the uncertainties associated with line outages and the influence of three different types of photovoltaics (PV) systems. The detailed transport network model capturing road congestion impacts is also missing in \cite{lei2019resilient,ye2020resilient,ding2020multiperiod,zhang2021stochastic}. In \cite{wang2021multi}, both MPSs and RCs are employed for the load restoration problem of an integrated power and hydrogen distribution network, where transport network and road congestion are captured.

Furthermore, the network reconfiguration of distribution networks is regarded as an efficient method and should be considered in the load restoration process \cite{ding2022sequential,ding2017new,ding2017resilient,wang2021multi}. In \cite{ding2017new,ding2017resilient}, a time-efficient network reconfiguration model is proposed for the resilience enhancement of distribution networks. A virtual network with a set of radiality constraints is firstly introduced in \cite{ding2017new}, which can significantly improve the computational performance. In \cite{ding2017resilient}, a network reconfiguration strategy considering microgrids (MGs), substations, and unsupplied load islands is developed for black-start load restoration after extreme events. In \cite{wang2021multi}, the multi-period network reconfiguration process is considered in an integrated power and hydrogen distribution network. To simulate more realistic load restoration process, a sequential black-start restoration model is proposed in \cite{ding2022sequential}, which can effectively capture the influence of black-start resources and dynamically concerns the sequence of recovered lines and buses.

Although the above model-based centralized optimization methods have been successfully applied to solve various MPS and RC resilience enhancement problems, the following challenges have to be addressed in the real-world environment. First, centralized optimization methods solely depend on a single control center, requiring prohibitive communication and computation resources. Furthermore, the centralized manner can be prone to single-point failure as all decisions are taken by this central controller \cite{li2019full,ge2022resilience}. Second, since power systems are moving towards a decentralized fashion, it is typically intractable to acquire operation models and technical parameters of MPSs and RCs. Third, because the power and transport networks are highly dynamic and stochastic, it is hard to generalize an adaptive control scheme that accounts for various system uncertainties (e.g., renewables, demand, traffic volumes, etc.). As the system uncertainties can be characterized by many factors (e.g., weather conditions, energy usage behaviors, driving habits, etc.), it is difficult to even represent the uncertainty probability distributions exactly. Last but not least, even though the system models and uncertainties can be known, solving a scenario-based stochastic optimization problem for load restoration is normally time-consuming. Therefore, developing a control scheme for these decentralized MPSs and RCs becomes important and urgent \cite{lei2018routing}.

\textit{Reinforcement learning} (RL) \cite{sutton2018reinforcement}, a data-driven and model-free method, may solve time-coupled decision-making problems by learning optimal policies through repeated interactions with the environment without any \textit{prior} knowledge. As an online learning method, RL can effectively utilize the growing amount of data from the environment, capture various uncertainties, and adjust to different state conditions. Furthermore, well-trained policies can be directly deployed to the practical test process in milliseconds without solving an optimization problem. Recently, RL has been successfully applied to many resilient power system operation problems, such as load restoration \cite{bedoya2021distribution,zhao2022deep}, voltage regulation \cite{kamruzzaman2021deep}, etc. However, the up-to-date literature investigating the application of RL to the dispatch problems of mobile resources towards MG load restoration is still limited. In \cite{yao2020resilient}, a single-agent reinforcement learning (SARL) method is applied to optimize the dispatch decisions of four MESSs for critical load restoration in MGs. However, applying SARL to a multi-agent setup may raise the scalability issue since the action space increases significantly with agent size, thereby costing computational time with one central agent \cite{zhang2020multi}. To address this issue, the authors in \cite{yi2022multi} propose a multi-agent reinforcement learning (MARL) method that models each MESS as an individual agent for MG load restoration via a multi-agent formulation. However, there are still several difficulties that can be identified and are worthy of further efforts. First, the coordinated effect of MPSs and RCS is not investigated in the above paper. It is noted that effectively coordinating these decentralized MPSs and RCs is capable of achieving better MG load restoration performance. Second, at each time step, the routing and scheduling actions are computed simultaneously. However, MPSs and RCs cannot simultaneously make routing decisions in the transport network and scheduling decisions in the power network because the decision-making processes in the two networks are mutually exclusive. Therefore, developing an effective MARL-based method to separately compute transport routing action and power scheduling action is important for MPSs and RCs to enhance MG resilience.    

\subsection{Contributions}
\label{sec:I.C}
Based on the review of previous work on both model-based \cite{arif2018optimizing,9330590,7559799,9121323,lei2018routing,yang2020seismic,lei2019resilient,ye2020resilient,ding2020multiperiod} and model-free \cite{bedoya2021distribution,zhao2022deep,kamruzzaman2021deep,yao2020resilient,yi2022multi} methods, a significant research gap has been identified, which drives the motivation behind this paper: no previous work has developed a coordinated control scheme of MPSs and RCs, operating in a decentralized manner for MG load restoration, and employing a model-free decision-making framework at the same time. To fill this knowledge gap, following contributions are achieved:  

\begin{enumerate}[label=\arabic*)]
\item Develop a novel decentralized framework for the coordinated decision-making problem of MPSs and RCs towards MG resilience enhancement with the objective of maximizing restored loads. This framework overcomes certain limitations of previous work: i) in contrast to \cite{lei2019resilient,ye2020resilient,ding2020multiperiod} neglecting the tranport sectors of mobile sources, the detailed transport network model capturing road congestion impacts is accounted for; and ii) in contrast to \cite{arif2018optimizing,9330590,7559799,9121323,lei2018routing,yang2020seismic} formulating the dispatch problem in a centralized manner, the decentralized co-dispatching behaviors of MPSs and RCs are captured that do not depend on the central commands, therefore avoiding the single-point failure and constructing the privacy perseverance.

\item Formulate the coordinated decision-making problem of MPSs and RCs as a \textit{Decentralized Partially Observable Markov Decision Process} (Dec-POMDP), wherein MPSs and RCs are regarded as the agents who can operate in a decentralized manner without following the central commands. In this context, the mobility and flexibility of MPSs and RCs can be fully explored inside the power-transport network towards system load restoration, in the meantime without knowing system models and uncertainty parameters.

\item Propose a novel hierarchical and hybrid MARL method to solve the Dec-POMDP. First, it learns a high-level (HL) policy that can direct MPSs and RCs in choosing between transport routing and power scheduling. Second, it learns a low-level (LL) hybrid policy that can capture MPSs' discrete routing and continuous scheduling actions. Additionally, RCs learn their routing and repairing selections in the HL via the same hierarchical structure, while learning their discrete routing and repairing actions in the LL via a conventional categorical policy. In this setting, the MPS and RC agents can both benefit from the hierarchical structure to learn the effective routing and scheduling/repairing actions separately rather than learning these two kinds of actions simultaneously, of which one becomes invalid. To further improve the scalability and stability of MARL policy, an abstracted embedded function capturing system dynamics is introduced during the training process.

\item Validate the superior performance of the proposed MARL method over the state-of-the-art model-based and model-free methods in MG load restoration. A generalized dispatch policy can be formulated and adapted to different sizes of MPSs and RCs and power-transport networks as well as system uncertainties.
\end{enumerate}

The rest of the paper is organized as follows. Section \ref{sec:II} presents the general formulations of the utilized MPSs and RCs as well as the operational models of both transport and power networks. Sections \ref{sec:III} and \ref{sec:IV} introduce the Dec-POMDP formulation and the proposed H2MAPPO method, respectively. In Section \ref{sec:V}, case studies are carried out and discussed on two experimental environments. Section \ref{sec:VI} draws the conclusions and future work of this paper.

\section{General Formulations of MPSs and RCs in Transport and Power Networks}
\label{sec:II}
\subsection{Problem Descriptions}
\label{sec:II.A}
\begin{figure}[t!] 
\centering  
\includegraphics[width=0.485\textwidth]{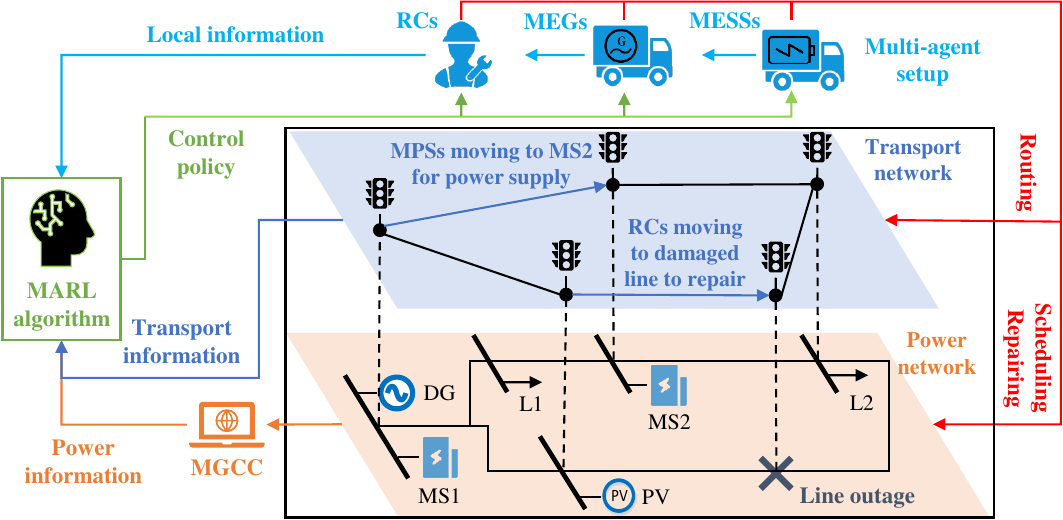}
\caption{The scheme of coordinated dispatch problem of MPSs and RCs in a power-transport network.}
\label{fig:scheme}
\end{figure}

We focus on the resilience-driven coordinated dispatch problem of MPSs and RCs within a power-transport network including both routing and scheduling/repairing behaviors, as illustrated in Fig.~\ref{fig:scheme}. In general, electric components (e.g., buses and lines) in the power network are located on different transport nodes, while MPSs and RCs can move upon the transport network and choose to connect with their corresponding candidate nodes \cite{7559799}. Specifically, we consider MEGs and MESSs as two types of MPSs that can choose to connect with the candidate nodes, e.g., MESS stations (MSs) \cite{lei2018routing}. Following \cite{zhang2021stochastic,lei2018routing}, this paper assumes that both MESSs and MEGs have black-start capability during the load restoration process. The role difference between MESSs and MEGs is that MESSs can charge power at one location with sufficient power supply and then discharge power at another location suffering load shedding, while MEGs can only provide power supply for the power network. In other words, MESSs play a role similar to demand-side response, whereas MEGs play a role similar to traditional generators but with mobility features. Regarding RCs, the candidate nodes are the initial depots and the locations of line outages. Inside the power network, static DERs, such as photovoltaics (PVs) and diesel generators (DGs), are installed suitably. In terms of the demand side, the power system captures both essential and non-essential loads to highlight the primary objective of load restoration \cite{wang2021three}. 

Unlike the centralized framework \cite{arif2018optimizing,9330590,7559799,9121323,lei2018routing,yang2020seismic,lei2019resilient,ye2020resilient,ding2020multiperiod,yao2020resilient}, this paper assumes that MPSs and RCs are operating in a decentralized manner, in anticipation of a future trend towards resilient distribution networks \cite{lei2018routing,wang2015research,li2019full,ge2022resilience}. In other words, MPSs and RCs can determine their individual dispatch behaviors without the central commands. In more detail, each resource can only acquire the local information of the coupled networks (e.g., PV generation, nodal load, line outage, and traffic volume) and its own status (e.g., transport location, battery state-of-charge (SoC), and repair capacity). Then, each resource determines its routing and scheduling/repairing decisions via an automatic control scheme. To solve the above decentralized coordination problem of MPSs and RCs, we first formulate this multi-agent setup as a Dec-POMDP, wherein MPSs and RCs are regarded as the agents and the coupled power-transport network is the environment. Then, a novel MARL method is proposed to drive MPS and RC agents to make optimal routing and scheduling/repairing actions in the transport and power networks respectively, aiming to maximize system resilience.

When the locations and repairs or power schedules of all mobile resources are settled, the \textit{microgrid central controller} (MGCC) regulates each controlled DER and smart switches optimally in the power network for weighted load restoration maximization towards resilience enhancement. It is worth noting that the dispatch decisions of MPSs and RCs and the smart switch operations for network reconfiguration are mutually influenced. On one hand, the different locations and dispatch behaviors of MPSs and RCs can lead to different network reconfiguration results; on the other hand, MPS and RC agents repeatedly interact with the power network environment during the RL training process, gradually learn the key features of the environment (e.g., DER schedules and smart switch locations), and then adjust their dispatch behaviors towards better load restoration performance. In this section, we present the general formulations of the routing behaviors in the transport network and the scheduling/repairing behaviors in the power network.

\subsection{Routing in Transport Network}
\label{sec:II.B}
\begin{figure}[h!] 
\centering  
\includegraphics[width=0.395\textwidth]{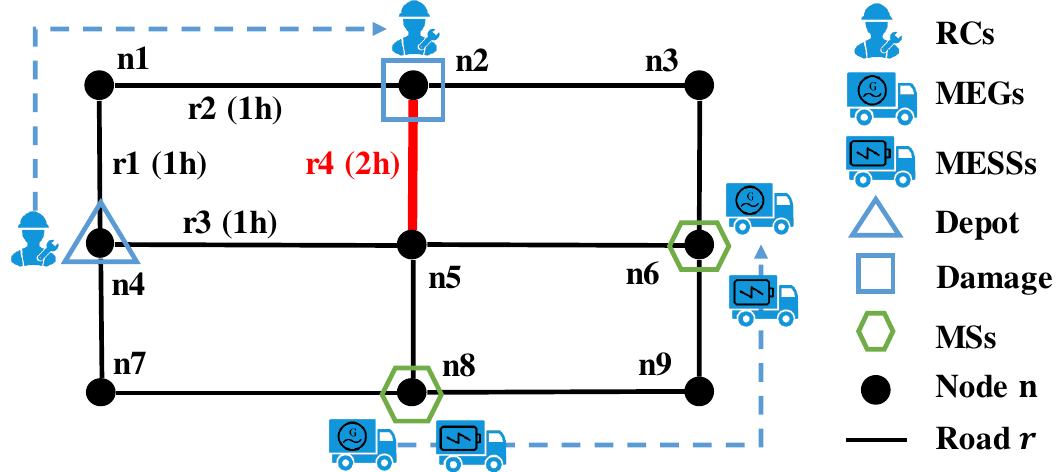}
\caption{Routing behaviors of MPSs and RCs in a transport network.}
\label{fig:route}
\end{figure}

The transport network is modeled as a weighted graph $G=\{N,R\}$, where $n \in N$ is the node set and $r \in R$ denotes the road set with the commuting distance $L_{r}$ \cite{yao2020resilient}. The general framework of MPSs and RCs' routing behaviors is illustrated in Fig. \ref{fig:route}. Specifically, to commute in the transport network, they need to consider the influence of uncertain travel time and rely on an effective routing model, which are described in the following subsections.

\subsubsection{Uncertain Travel Time}
\label{sec:II.B.1}
The travel time $T_{r,t}^{rd}$ of road $r$ is influenced by the real-time traffic volume \cite{yuanqing2004theory}, which can be estimated as
\begin{equation}\label{eq:ev flow1}
    T_{r,t}^{rd} = \tilde{T}_{r}^{rd}[1+\alpha^{rd}(\frac{V^{rd}_{r,t}}{C_{r}})^{\beta^{rd}}], \forall r \in R, \forall t \in T,
\end{equation}
where $\tilde{T}_{r}^{rd}$ is the free driving time of road $r$ that mainly depends on the road distance, while $C_{r}$, $\alpha^{rd}$, $\beta^{rd}$, and $V^{rd}_{r,t}$ correspond to the capacity, the retardation coefficients and the traffic volume of road $r$ at time step $t$, respectively. Equation \eqref{eq:ev flow1} can characterize the relationship between uncertain traffic volume and travel time, as well as reflect road impedance based on traffic flow itself. Note that heavy traffic volumes can cause serious road congestion and long travel time. It is crucial to account for the impact of road congestion on realistic routing behaviors; nevertheless, most existing literature on MPS and RC routing problems (e.g., \cite{lei2019resilient,ye2020resilient,ding2020multiperiod}) ignores this factor. It is worth noting that the travel time between candidate nodes is directly associated with the dispatch behaviors of MPSs and RCs, while serious road congestion can significantly increase the required travel time. As such, ignoring road congestion impact can lead to inaccurate estimation of travel time and impractical dispatch behaviors of MPSs and RCs.

\subsubsection{Routing Model}
\label{sec:II.B.2}
The routing behaviors of mobile units $i \in I$ including both MPSs and RCs within the transport network $G$ can be formulated in a same manner \cite{lei2019resilient}, which are restricted by
\begin{equation}\label{eq:route_1}
    \sum_{n \in N} u_{i,n,t} \leq 1, \forall i \in I, \forall t \in T,
\end{equation}
\begin{equation}\label{eq:route_2}
\begin{split}
    \!\!\sum_{\tau = t}^{\min (t+T^{rd}_{mn,t},T)} \!\!u_{i,m,\tau} \leq (1-u_{i,n,t}) \cdot \min(T^{rd}_{mn,t}, T-t), \\ \forall i \in I, \forall n \in N, \forall m \in N\backslash\{i\}, \forall t \in T.
\end{split}
\end{equation}
where constraint~\eqref{eq:route_1} shows that a mobile unit $i$ can only be connected with one transport node for each time step, the binary variable $u_{i,n,t} \in \{0,1\}$ indicates the the mobile unit $i$ connecting to the node $n$ ($u_{i,n,t}=1$) or not ($u_{i,n,t}=0$) at time step $t$. Constraint \eqref{eq:route_2} ensures reasonable routing behaviors of the mobile unit $i$ between different transport nodes, where $T^{rd}_{mn,t}$ refers to the time period required to route from node $n$ to node $m$ at time step $t$. Note that $T^{rd}_{mn,t}$ can be uncertain and influenced by the real-time traffic volume $V^{rd}_{mn,t}$. 

\subsubsection{Routing Behaviors}
\label{sec:II.B.3}
In order to clearly illustrate the routing behaviors simulated in this paper, we take RCs in Fig. \ref{fig:route} as an example, the depot location of RCs is $n_{4}$ and the location of damaged line is $n_{2}$. Inside the network topology $G$, there are two available routes (i.e., $r_1 \rightarrow r_2$ and $r_3 \rightarrow r_4$) for RCs to commute, while serious congestion happening on road $r_4$ (the red road in Fig. \ref{fig:route}) leads to much longer travel time ($T_{r_{3+4},t}^{rd}=3$ hours) of route $r_3 \rightarrow r_4$. In this context, RCs will choose $r_1 \rightarrow r_2$ as their commuting route rather than $r_3 \rightarrow r_4$ due to the less 1 hour travel time. Regarding MEGs and MESSs, their routing behaviors in the network topology $G$ can be derived in the similar manner as RCs. As shown in Fig. \ref{fig:route}, the examined MEGs and MESSs are traveling from the initial MS at $n_{8}$ to the MS at $n_{6}$. Finally, when MPSs and RCs arrive at their destinations, they can be connected to the power network via MSs and damaged components, represented by $N_{ms}$ and $N_{rc}$ as subsets of $N$ \cite{lei2019resilient}.

\subsection{Scheduling/Repairing in Power Network}
\label{sec:II.C}
In power network, MPSs make scheduling decisions (i.e., MEG $g$ power output $P_{g,n,t}^{eg}$, MESS $k$ charging power $P_{k,n,t}^{esc}$ and discharging power $P_{k,n,t}^{esd}$) in MS $n$, RC $j$ makes repairing decision $Re_{j,w,t}^{rc}$ for damaged component $w$. Afterwards, a linearized alternating current optimal power flow (AC-OPF) algorithm with the objective of load restoration maximization towards MG resilience enhancement can be solved by the MGCC for each time step $t$. The operation models of MEG, MESS, and RC as well as the employed AC OPF algorithm can be found in the following subsections.

\subsubsection{MEG Scheduling}
\label{sec:II.C.1}
MEGs are constrained by their active and reactive power output limits
\begin{equation} \label{eq:meg_1}
    \underline{P}^{eg}_{g} \leq P^{eg}_{g,n,t} \leq \overline{P}^{eg}_{g}, \forall g \in I_{eg}, \forall n \in N_{ms}, \forall t \in T,
\end{equation}
\begin{equation} \label{eq:meg_2}
    \underline{Q}^{eg}_{g} \leq Q^{eg}_{g,n,t} \leq \overline{Q}^{eg}_{g}, \forall g \in I_{eg}, \forall n \in N_{ms}, \forall t \in T.
\end{equation}
where $P^{eg}_{g,n,t}$ and $Q^{eg}_{g,n,t}$ correspond to active and reactive power output of MEG $g$ in MS $n$ at time step $t$, respectively. We assume that adequate fuel is available for MPS routing through portable or towable fuel tanks and optimally dispatched fuel trucks in case of long term blackouts, following \cite{lei2019resilient,lei2018routing,zhang2021stochastic,yang2020seismic,7559799}. Additionally, to extend the continuous operating time of MPSs and handle the potential fuel issues, there are several potential solutions, e.g., appropriately selecting candidate nodes to connect MPSs and fuel trucks, deploying underground fuel tanks at candidate nodes, and pre-allocating fuel in the network \cite{lei2019resilient,7559799}. As such, the detailed logistic process based on the location of the re-filling station and the optimal re-filling time is not considered in this paper.

\subsubsection{MESS Scheduling}
\label{sec:II.C.2}
The power operation model of MESSs can be appropriately presented via
\begin{equation}\label{eq:mess_1}
    0 \leq P^{esc}_{k,n,t} \leq u_{k,t}^{es} \overline{P}_{k}^{es}, \forall k \in I_{es}, \forall n \in N_{ms}, \forall t \in T,
\end{equation}
\begin{equation}\label{eq:mess_2}
\!\!\!-\overline{P}_{k}^{es} (1-u_{k,t}^{es}) \leq P^{esd}_{k,n,t} \leq 0, \forall k \in I_{es}, \forall n \in N_{ms}, \forall t \in T,
\end{equation}
\begin{equation}\label{eq:mess_3}
    \underline{S}^{es}_{k} \leq S^{es}_{k,t} \leq \overline{S}^{es}_{k}, \forall k \in I_{es}, \forall t \in T,
\end{equation}
\begin{equation} \label{eq:mess_4} 
\begin{split}
    S_{k,t+1}^{es} = S_{k,t}^{es} + \frac{P^{esc}_{k,n,t}\eta_{k}^{esc} + P^{esd}_{k,n,t}/\eta_{k}^{esd}}{\overline{E}_{k}^{es}}, \\ \forall k \in I_{es}, \forall n \in N_{ms}, \forall t \in T,
\end{split}
\end{equation}
where constraints~\eqref{eq:mess_1} and \eqref{eq:mess_2} restrict the maximum charging and discharging power of MESS $k$ in MS $n$ at time step $t$. The binary variable $u_{k,t}^{es} \in \{0,1\}$ introduced in constraints~\eqref{eq:mess_1} and \eqref{eq:mess_2} indicates the charging ($u_{k,t}^{es}=1$) or discharging ($u_{k,t}^{es}=0$) behavior of MESS $k$ at time step $t$. It is noted that these two behaviors cannot happen simultaneously. Constraint \eqref{eq:mess_3} limits the minimum and maximum battery SoC level of MESS $k$, while its dynamic transition between two consecutive time steps is presented in \eqref{eq:mess_4}, given the charging/discharging power $P^{esc}_{k,t},P^{esd}_{k,t}$ and efficiencies $\eta_{k}^{esc},\eta_{k}^{esd}$.

\subsubsection{RC Repairing}
\label{sec:II.C.3}
The repair plan of RC $j$ is formulated as
\begin{equation}\label{eq:rc_3}
    z^{rc}_{j,w,t} \leq \frac{\sum_{\tau=1}^{t} Re^{rc}_{j,w,\tau}}{RT_{w}^{rc}}, \forall j \in I_{rc}, \forall w \in N_{rc}, \forall t \in T,
\end{equation}
\begin{equation}\label{eq:rc_4}
    z^{rc}_{j,w,t} \leq z^{rc}_{j,w,t+1}, \forall j \in I_{rc}, \forall w \in N_{rc}, \forall t \leq T-1,
\end{equation}
\begin{equation}\label{eq:rc_5}
    \sum_{w \in N_{rc}} rs^{rc}_{w} \cdot z^{rc}_{j,w,T} \leq RS^{rc}_{j}, \forall j \in I_{rc}.
\end{equation}
where the binary variable $z^{rc}_{j,w,t}=1$ if the damaged component $w$ is repaired by RC $j$ at time step $t$, and $z^{rc}_{j,w,t}=0$ otherwise. Binary $Re^{rc}_{j,w,t}$ represents if RC $j$ is repairing component $w$ at time step $t$ (1 if repairing, 0 otherwise), while $RT_{w}^{rc}$ corresponds to the time period required to repair component $w$ \cite{lei2019resilient}. Constraint \eqref{eq:rc_5} ensures that the resource capacity $RS^{rc}_{j}$ of RC $j$ is sufficient for its repair tasks, where $rs^{rc}_{w}$ refers to the number of resources required to repair damaged component $w$.

\subsubsection{Power Network}
\label{sec:II.C.4}
The power network operation is fully modeled by a linearized OPF algorithm for MG resilience enhancement. Specifically, once MEG $g$, MESS $k$ and RC $j$ are moving to their individual destinations, where MEG $g$ and MESS $k$ are making their power schedules $P_{g,t}^{eg}$ and $P_{k,t}^{es}$, while RC $j$ is making its repairing decision $Re_{j,w,t}^{rc}$ for damaged component $w$. The following linearized OPF algorithm can be then solved by the MGCC for each time step $t$.
\begin{equation}\label{eq:obj}
   \Big\{ \max_{\Xi^{mg}} \mathbb{E} \big\{\sum_{d \in D} c_{d}^{ls} P_{d,t}^{ed} \big\},
\end{equation}
\text{where}
\begin{equation}\label{eq:acopf_var}
\begin{split}
    \Xi^{mg} = \{P_{d,t}^{ed}, Q_{d,t}^{ed}, P_{g,t}^{dg}, Q_{g,t}^{dg}, P^{pv}_{g,t}, Q_{g,t}^{pv}, \\ P_{bp,t}, Q_{bp,t}, V^2_{b,t}, y_{bp,t},e_{b,t}, F_{bp,t}\},
\end{split}
\end{equation}
\text{subject to}
\begin{equation}\label{eq:api}
\begin{split}
    \! \sum_{g \in B_{dg}}\!\!P_{g,t}^{dg} +\! \sum_{g \in B_{eg}}\!\!P_{g,t}^{eg}+\! \sum_{g \in B_{pv}}\!\!P_{g,t}^{pv} =\!\! \sum_{d \in B_{ed}}\!\!P^{ed}_{d,t} \\+\! \sum_{k \in B_{es}}\!\!P_{k,t}^{es}-\! \sum_{(p,b) \in L}\!\!P_{pb,t} +\! \sum_{(b,p) \in L}\!\!P_{bp,t}, ~\forall b \in B,
\end{split}
\end{equation}
\begin{equation}\label{eq:rpi}
\begin{split}
    \sum_{g \in B_{dg}}Q_{g,t}^{dg}+\sum_{g \in B_{eg}}\!\!Q_{g,t}^{eg}+\! \sum_{g \in B_{pv}}\!\!Q_{g,t}^{pv} = \sum_{d \in B_{ed}}Q^{ed}_{d,t}\\-\sum_{(p,b) \in L}Q_{pb,t} + \sum_{(b,p) \in L}Q_{bp,t}, ~\forall b \in B,
\end{split}
\end{equation}
\begin{equation} \label{eq:adg}
    \underline{P}^{dg}_{g} \leq P^{dg}_{g,t} \leq \overline{P}^{dg}_{g}, ~\forall g \in DG,
\end{equation}
\begin{equation} \label{eq:rdg}
    \underline{Q}^{dg}_{g} \leq Q^{dg}_{g,t} \leq \overline{Q}^{dg}_{g}, ~\forall g \in DG,
\end{equation}
\begin{equation} \label{eq:b_pv}
    0 \leq P_{g,t}^{pv} \leq \tilde{P}^{pv}_{g,t},~\forall g \in PV^{bs},
\end{equation}
\begin{equation}\label{eq:b_pv_s}
    (P^{pv}_{g,t})^2+(Q^{pv}_{g,t})^2 \leq (\overline{S}^{pv}_{g})^{2},~\forall g \in PV^{bs},
\end{equation}
\begin{equation} \label{eq:n_pv}
    0 \leq P_{g,t}^{pv} \leq e_{b,t}\tilde{P}^{pv}_{g,t},~\forall b \in B,~\forall g \in B_{pv} \cap PV^{nbs},
\end{equation}
\begin{equation}\label{eq:n_pv_s}
   \! (P^{pv}_{g,t})^2+(Q^{pv}_{g,t})^2 \leq e_{b,t}(\overline{S}^{pv}_{g})^{2},\forall b \in B,\forall g \in B_{pv} \cap PV^{nbs},
\end{equation}
\begin{equation} \label{eq:energize_load}
P^{ed}_{d,t} \leq e_{b,t} \overline{P}^{ed}_{d,t},~\forall b \in B,\forall d \in B_{ed},
\end{equation}
\begin{equation} \label{eq:energize_load_q}
Q^{ed}_{d,t} \leq e_{b,t} \overline{Q}^{ed}_{d,t},~\forall b \in B,\forall d \in B_{ed},
\end{equation}
\begin{equation}\label{eq:vol}
    e_{b,t} \underline{V}^2 \leq V^2_{b,t} \leq e_{b,t} \overline{V}^2, ~\forall b \in B,
\end{equation}
\begin{equation}\label{eq:thermal}
    P_{bp,t}^2 + Q_{bp,t}^2 \leq y_{bp,t} \cdot \overline{S}_{bp}, ~\forall (b,p) \in L,
\end{equation}
\begin{equation}\label{eq:apf}
\begin{split}
    V^2_{b,t} - V^2_{p,t} \leq 2 \cdot (r_{bp} P_{bp,t} + x_{bp} Q_{bp,t}) \\ + (1-y_{bp,t}) \cdot M, ~\forall (b,p) \in L,
\end{split}
\end{equation}
\begin{equation}\label{eq:rpf}
\begin{split}
    V^2_{b,t} - V^2_{p,t} \geq 2 \cdot (r_{bp} P_{bp,t} + x_{bp} Q_{bp,t}) \\+ (y_{bp,t}-1) \cdot M, ~\forall (b,p) \in L,
\end{split}
\end{equation}
\begin{equation} \label{eq:radial_1}
    \sum_{(b,p) \in L}y_{bp,t} = |B|-\big( \sum_{b \in B}(1-e_{b,t})+N_{bs} \big),
\end{equation}
\begin{equation} \label{eq:radial_2}
\sum_{a \in B_{bs}}F^{s}_{a,t}+\sum_{(p,b) \in L} F_{pb,t} - \sum_{(b,p) \in L} F_{bp,t} = e_{b,t}, ~\forall b \in B,
\end{equation}
\begin{equation} \label{eq:radial_3}
- y_{bp,t} \overline{F}_{bp} \leq F_{bp,t} \leq y_{bp,t} \overline{F}_{bp}, ~\forall (b,p) \in L,
\end{equation}
\begin{equation} \label{eq:radial_4}
    y_{bp,t} \leq z_{j,bp,t}^{rc}, ~\forall (b,p) \in N_{rc}, ~\Big\}, \quad t \in T,
\end{equation}
where the objective function \eqref{eq:obj} is to maximize the expectation of weighted sum of restored loads \cite{lei2019resilient,ye2020resilient,ding2020multiperiod} capturing various uncertainties (e.g., demand and PV generation) and stochastic variables in set $\Xi^{mg}$. The OPF constraints formulated by the LinDistFlow model \cite{baran1989network} include the active and reactive power balances \eqref{eq:api}-\eqref{eq:rpi} at bus $b$, where $B_{eg}$, $B_{es}$, $B_{ed}$, $B_{dg}$, and $B_{pv}$ correspond to the sets of MEG $g$, MESS $k$, restored load $d$, DG $g$ and PV $g$ located at bus $b$, respectively. The active and reactive power outputs of DG $g$ are constrained by \eqref{eq:adg} and \eqref{eq:rdg} respectively, while the output limits of grid-forming and grid-following PVs are presented in \eqref{eq:b_pv}-\eqref{eq:b_pv_s} and \eqref{eq:n_pv}-\eqref{eq:n_pv_s} respectively \cite{zhang2021two}. The status of load $d$ is restricted by constraint \eqref{eq:energize_load} and \eqref{eq:energize_load_q}, where binary $e_{b,t}$ corresponds to the energized status of bus $b$ (1 if energized, 0 otherwise). The voltage and power flow limits are shown in \eqref{eq:vol} and \eqref{eq:thermal} respectively, while the linearized power flow constraints are expressed in \eqref{eq:apf}-\eqref{eq:rpf}. Binary $y_{bp,t}$ in \eqref{eq:thermal}-\eqref{eq:rpf} indicates the energized status of line $(b,p)$ (1 if energized, 0 otherwise) determined by both reconfiguration switch and RC operations, while $M$ refers to a large positive number used to relax constraints \eqref{eq:apf}-\eqref{eq:rpf} for disconnected or damaged lines. 

To coordinate with mobile source dispatch, the power network can be dynamically reconfigured through smart switch operations, while this process should respect the system radiality, subject to a set of virtual network constraints \eqref{eq:radial_1}-\eqref{eq:radial_4} \cite{ding2022sequential,zhang2021stochastic}. 
According to \cite{ding2017new,ding2017resilient,wang2021multi}, two conditions should be satisfied for the network radiality: i) each island is connected; ii) the number of energized lines is equal to the number of buses minus the number of islands. Furthermore, an additional condition is introduced by \cite{ding2022sequential} to ensure that de-energized or cut-off buses are not connected with each other due to the lack of black-start resources.
Specifically, constraint \eqref{eq:radial_1} maintains the network radiality during the network reconfiguration process, which shows that the number of lines is equal to the number of buses minus the number of possible islands containing black-start resources and de-energized buses \cite{ding2022sequential}. Constraint \eqref{eq:radial_2} refers to the nodal power balance of the virtual network, indicating that bus $b$ is energized if the virtual load at the bus is served ($x_{b,t}=1$). $B_{bs}$ is the set of power sources with black-start capabilities (e.g., buses connected with DGs, grid-forming PVs, and MPSs in the real power network) located at bus $b$. Constraint \eqref{eq:radial_3} models the connections between the power network and the virtual network through the line connection status $y_{bp,t}$. Constraint \eqref{eq:radial_4} indicates that the damaged line $(b,p) \in N_{rc}$ could be energized in the power network after being repaired by RC $j$ ($z_{j,bp,t}^{rc}=1$). More detailed mathematical formulations of the power network model can be found in \cite{lei2019resilient,ding2022sequential,ding2017new,ding2017resilient,zhang2021stochastic}.

\section{Reformulation as a Dec-POMDP}
\label{sec:III}
The conventional method lies in using optimization approach to solve above coordinated dispatch problem of MPSs and RCs \cite{lei2019resilient,ye2020resilient,ding2020multiperiod}. However, it is hard to generalize an adaptive and fast control scheme that accounts for various system dynamics and uncertainties in the context of power and transport networks, as discussed in Section \ref{sec:I}. To this end, it is reasonable to formulate this problem as a \textit{Decentralized Partially Observable Markov Decision Process} (Dec-POMDP) \cite{oliehoek2016concise}, considering that MPSs and RCs are operating in a decentralized manner and can only observe partial information of the power and transport networks.

A Dec-POMDP is a 7-tuple $\langle \mathcal{I}, \mathcal{S}, \{\mathcal{O}_{i}\}, \{\mathcal{A}_{i}\}, \mathcal{R}, \mathcal{T}, \gamma \rangle$, including $|\mathcal{I}|$ agents, a collection of global states $s \in \mathcal{S}$, local observations $\{o_{i} \in \mathcal{O}_{1:I}\}$, action sets $\{a_{i} \in \mathcal{A}_{1:I}\}$, and reward functions $\{r_{i} \in \mathcal{R}\}$, as well as a state transition function $\mathcal{T}(s,o_{1:I},a_{1:I},\omega)$, where $\omega$ is the environment stochasticity representing the system uncertain parameters. The time interval $\Delta t$ is 1 hour. For each agent $i$ at time step $t$, an action $a_{i,t}$ is computed using the policy $\pi_{i}(a|o)$ conditioned on the current local observation $o_{i,t}$. Then, the environment transits to the next state given the transition function $\mathcal{T}$, while agent $i$ is rewarded $r_{i,t}$ and updated a next local observation $o_{i,t+1}$. Following this process, each agent $i$ receives a trajectory of local observations, actions, and rewards: $\tau_{i} = o_{i,1},a_{i,1},r_{i,1},o_{i,2},...,r_{i,T}$ over $\mathcal{O}_{i} \times \mathcal{A}_{i} \times \mathcal{R} \rightarrow \mathbb{R}$. The objective of each agent $i$ is maximizing its cumulative discounted reward $R_{i} = \sum^{T}_{t=0} \gamma^{t} r_{i,t}$, where $\gamma \in [0,1)$ and $T$ = 24 hours are the discount factor and daily horizon, respectively. The components of Dec-POMDP are as below:

\subsection{Agent}
The agents are defined as three groups of resources: 1) RC agent $i = j \in I_{rc} \subseteq I$; 2) MEG agent $i = g \in I_{eg} \subseteq I$; and 3) MESS agent $i = k \in I_{es} \subseteq I$. 

\subsection{Environment}
The environment can be organized into two sectors: 1) routing process of RC, MEG and MESS agents in the transport network; 2) repairing/scheduling process of RC, MEG and MESS agents in the power network. The routing constraints of mobile sources \eqref{eq:route_1} and \eqref{eq:route_2} can be satisfied automatically by realistic transport network settings (e.g., transport nodes, road distance, and traffic volumes) in the RL environment.

\subsection{Observation}
Each agent $i$ at time step $t$ observes its local observation $o_{i,t}$ differing for distinct groups, defined as
\begin{equation} \label{eq:observation} 
    \!o_{i,t}=
    \begin{cases}
    [N_{j,t}, V_{j,t}^{rd}, S_{j,t}^{ln}, RT_{j,t}^{rc}, RS_{j,t}^{rc}, S_{j,t}^{rc}] & \!\!\text{$\forall i = j \in I_{rc}$} \\
    [N_{g,t}, V_{g,t}^{rd}, S_{g,t}^{ln}, P_{g,t}^{ed}, P_{g,t}^{pv}] & \!\!\text{$\forall i = g \in I_{eg}$} \\
    [N_{k,t}, V_{k,t}^{rd}, S_{k,t}^{ln}, P_{k,t}^{ed}, P_{k,t}^{pv}, S_{k,t}^{es}] & \!\!\text{$\forall i = k \in I_{es}$}
    \end{cases},
\end{equation}
comprising two parts: 1) the transport node index $N_{i,t}$ the agent $i$ is traveling on and the corresponding road traffic volume $V_{i,t}^{rd}$; 2) the power information of the line status (outage or not) $S_{i,t}^{ln}$, nodal load $P_{i,t}^{ed}$, PV generation $P_{i,t}^{pv}$, the battery SoC $S^{es}_{i,t}$ of MESS agent $i$, the time $RT_{i,t}^{rc}$ and resources $RS_{i,t}^{rc}$ required to repair the damaged line as well as the current resource status $S_{i,t}^{rc}$ of RC agent $i$. 

This paper assumes that agents can get access to the above required local information through interactions with the environment. However, it is worth noting that there exists a risk that these agents can only get access to the incomplete knowledge (or even no information) of the distribution network due to damaged communication infrastructure or data privacy concerns \cite{zhang2019learning}. In this context, to discover accurate nodal information only using incomplete knowledge, the following two methods may be useful: 1) instead of directly observing the detailed nodal knowledge of the coupled system, agents can choose to acquire aggregated information (e.g., aggregated load profiles) or estimated information (e.g., solar irradiance) \cite{zhang2019learning}; 2) agents can be equipped with effective forecasting mechanisms (e.g., long short-term memory (LSTM) networks \cite{ruan2021estimating}), which takes the incomplete information as input and then output the required local information for the RL training process.

\subsection{Action}
Each agent $i$ at time step $t$ controls its action $a_{i,t}$ that varies for different agent groups, defined as
\begin{equation} \label{eq:action}
    a_{i,t}=
    \begin{cases}
    [a_{j,t}^{l,rc}, a_{j,t}^{r,rc}] & \text{$\forall i=j \in I_{rc}$} \\
    [a_{g,t}^{l,eg}, a_{g,t}^{p,eg}] & \text{$\forall i=g \in I_{eg}$} \\
    [a_{k,t}^{l,es}, a_{k,t}^{p,es}] & \text{$\forall i=k \in I_{es}$} \\
    \end{cases},
\end{equation}
comprising two parts: 1) discrete routing action $a_{i,t}^{l}\in\{0,1,...,R^{rd}\}$ is selected from the set of potential routes upon the transport node, in which 0 denotes no routing behaviors and $R^{rd}$ denotes the number of available commuting routes at current transport node $N_{i,t}$, as described in \cite{zhao2020hybrid,liang2022real,huang2022charging}; 2) discrete action $a_{j,t}^{r,rc} \in \{0,1\}$ corresponds to the choice of RC $j$ to repair the component ($a_{j,t}^{r,rc}=1$) or not ($a_{j,t}^{r,rc}=0$); 3) continuous actions $a_{g,t}^{p,eg} \in [0,1]$ and $a_{k,t}^{p,es} \in [-1,1]$ represent the magnitude of power generation (for MEG agent $g$) and power charging/discharging (for MESS agent $k$) as a percentage of their power capacity $[\underline{P}^{eg}_{g},\overline{P}^{eg}_{g}]$ and $[-\overline{P}_{k}^{es},\overline{P}_{k}^{es}]$, respectively. 

\subsection{State Transition}
The state transition process from time step $t$ to $t+1$ is governed by $s_{t+1}=\mathcal{T}(s_{t},o_{1:I,t},a_{1:I,t},\omega_{t})$, which is affected by a combination of the current state $s_{t}$ of environment, local observations $o_{1:I,t}$ and actions $a_{1:I,t}$ of agents, and environment stochasticity $\omega_{t}$. Regarding this problem, the environment stochasticity $\omega_{t}=[S_{l,t}^{ln},P_{d,t}^{ed},P_{g,t}^{pv},V_{r,t}^{rd}]$ corresponds to the exogenous states that are independent of agent actions and have intrinsic variability. RL can overcome this variability by adopting a data-driven fashion that does not depend on precise probability distributions for various uncertainties but instead learns state features from the data set itself \cite{sutton2018reinforcement}. To better prove the effectiveness of RL on handling environment uncertainties, a test dataset (separate from training dataset) is normally used to evaluate the performance of the trained RL policy on generalization to different state conditions. Furthermore, once the RL policy is well trained, it can be directly deployed to the practical test process in milliseconds.

On the other hand, the transition of endogenous states $N_{i,t},RT_{i,t}^{rc},RS_{i,t}^{rc},S_{i,t}^{rc},S_{i,t}^{es}$ can be determined by the agents' action $a_{i,t}$. Specifically, $N_{i,t}$ is determined by $a_{i,t}^{l}$, corresponding to the routing decisions upon the transport network, as expressed in Section \ref{sec:II.B}. Furthermore, as an RC agent $i$, once moving to a damaged line, the repairing time $RT_{i,t}^{rc}$ and resources $RS_{i,t}^{rc}$ of this line can be observed. If RC agent $i$ decides to repair this line ($a^{r,rc}_{i,t}=1$), the remaining resources $S_{i,t}^{rc}$ at time step $t$ can be updated after the line is repaired:
\begin{equation} \label{eq:rcs} 
    S_{i,t+1}^{rc}=S_{i,t}^{rc}- a^{r,rc}_{i,t} RS_{i,t}^{rc}, \forall i \in I_{rc}.
\end{equation}

Finally, as a storage unit, $S_{i,t}^{es}$ of MESS agent $i$ is managed by its continuous action $a_{i,t}^{p,es}$ through the mutually exclusive quantities $P_{i,t}^{esc},P_{i,t}^{esd}$, which are limited by its minimum and maximum SoC level $\underline{S}_{i}^{es},\overline{S}_{i}^{es}$, energy and power capacities $\overline{E}^{es}_{i},\overline{P}^{es}_{i}$, and charging/discharging efficiency $\eta_{i}^{es}$, depicted as
\begin{equation} \label{eq:meesc} 
    \!P_{i,t}^{esc}=[\textrm{min}(a_{i,t}^{p,es} \overline{P}_{i}^{es}, (\overline{S}_{i}^{es}-S_{i,t}^{es}) \overline{E}_{i}^{es} / \eta_{i}^{es}]^{+},\forall i \in I_{es},
\end{equation}
\begin{equation} \label{eq:meesd} 
    \!P_{i,t}^{esd}=[\textrm{max}(a_{i,t}^{p,es} \overline{P}_{i}^{es}, (\underline{S}_{i}^{es}-S_{i,t}^{es}) \overline{E}_{i}^{es} \eta_{i}^{es}]^{-},\forall i \in I_{es},
\end{equation}
where operator $[\cdot]^{+/-}=\max/\min\{\cdot,0\}$. Then, the state transition of $S_{i,t}^{es}$ from time step $t$ to $t+1$ is written as 
\begin{equation} \label{eq:messsoc} 
    \centering
    S_{i,t+1}^{es} = 
    \begin{cases}
    S_{i,t}^{es} + \frac{P^{esc}_{i,t}\eta_{i}^{es} + P^{esd}_{i,t}/\eta_{i}^{es}}{\overline{E}_{i}^{es}} & \text{if~}a^{l,es}_{i,t}=0 \\
    S_{i,t}^{es} & \text{otherwise}
    \end{cases},\forall i \in I_{es},
\end{equation}
where $a_{i,t}^{l,es}=0$ indicates that the MESS agent $i$ connects to the grid at time step $t$.

\subsection{Reward}
After the dispatches of all MPSs and RCs have been obtained, load restoration quantity $P_{d,t}^{ed}$ of each load $d$ at time step $t$ can be optimized via the linearized AC-OPF algorithm described in Section \ref{sec:II.C}. It is noted that all the constraints inside the distribution network can be satisfied by solving the optimization \eqref{eq:obj}-\eqref{eq:radial_4} with sufficient flexibility supported by DERs, load shedding, and switch switch operations. The studied problem aims at maximizing the weighted sum of restored loads for resilience enhancement. Thus, the reward function of agents in Dec-POMDP is designed as the following resilience index
\begin{equation} \label{eq:reward} 
    r_{i,t} = \lambda_{t} = \frac{\sum\nolimits_{d \in D} c^{ls}_{d}P_{d,t}^{ed}}{\sum\nolimits_{d \in D} c^{ls}_{d}\overline{P}^{ed}_{d,t}}, \forall i \in I,
\end{equation}
where the higher $\lambda_{t}$ indicates the more restoration of weighted loads and consequently the better performance of MG resilience enhancement. The designed reward function \eqref{eq:reward} is similar to the objective function \eqref{eq:obj}, where the only difference is that the total weighted baseline loads are added in \eqref{eq:reward} to realize reward signals as unitless scalar values \cite{sutton2018reinforcement}. In Section \ref{sec:II.C.4}, the objective of the problem is to maximize the weighted load restoration of the distribution network. However, directly using equation \eqref{eq:obj} as the reward function may raise serious convergence and optimality issues. This is because of the large fluctuations of weighted load restoration for different state conditions, possibly learning the unbalanced distributions of weight and bias values of the control policies \cite{sutton2018reinforcement}. To combat this, we scale the weighted load restoration and introduce the (unitless) resilience index $\lambda_{t} \in [0,1]$ as the reward function.

\section{Proposed MARL Method}
\label{sec:IV}
\begin{figure}[t!]
\centering
\includegraphics[width=0.485\textwidth]{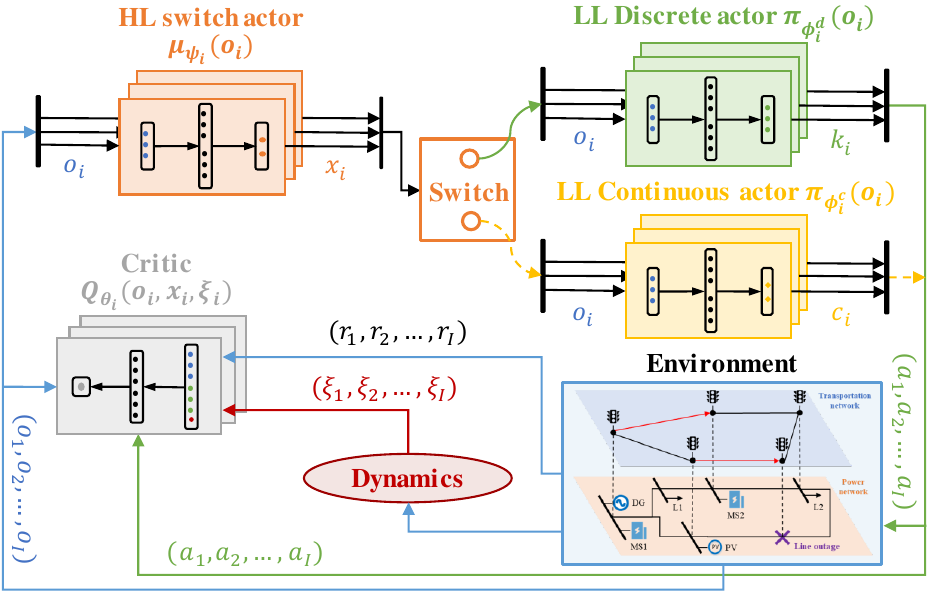}
\caption{The structure of the proposed H2MAPPO method.}
\label{fig:h2mappo}
\end{figure}

In this section, a MARL method called H2MAPPO is proposed to solve the above Dec-POMDP, with the overall architecture depicted in Fig.~\ref{fig:h2mappo}. Specifically, H2MAPPO generates four practical implementation details that are insightful and crucial: 1) constructing a hierarchical architecture using a two-level framework \cite{qiu2022multi} to choose between transport network routing and power network scheduling/repairing; 2) creating a hybrid policy \cite{qiu2022hybrid} that can perform both discrete routing and continuous scheduling actions; 3) updating the MARL policy using MAPPO algorithm \cite{yu2021surprising} that exhibits a stable learning performance, and is easy to implement, sample, and tune hyperparameters; 4) utilizing an embedded function to encapsulate system dynamics and approximate an abstracted state-value function, which can enhance the multi-agent training performance while providing privacy protection.

\subsection{Learn Two-Level Hierarchies}
\label{sec:IV.A}
Hierarchical reinforcement learning (HRL) refers to a type of RL method that can deal with several sub-policies working together in a hierarchical structure \cite{sutton1999between}. The two-level framework \cite{qiu2022multi}, one of the most common HRL techniques, is proposed as a temporal abstraction for RL actions, where the high-level (HL) action takes place over several time steps via the low-level (LL) actions. Specifically, for any RC or MPS agent $i$ observing $o_{i,t}$ at time step $t$, an HL action is chosen using the HL policy $x_{i,t} = \mu(x|o) \rightarrow [0,1] \in \mathcal{X}_{i}$ (e.g., when MESS $i$ is parked to a MS at time step $t$, it has a $x_{i,t}=80\%$ probability to choose charging/discharging actions in the power network and the left 20\% probability to choose routing actions in the transport network, then the final HL action would be choosing charging/discharging behaviors in the power network, this is because final selection would be the one with the higher sampling probability). Afterwards, the LL policy $\pi(a|o)$ is utilized to compute the LL action $a_{i,t}$ (e.g., discharge MESS battery power to support load restorations). This process continues until the HL action switches to the transport network routing when probability $x_{i,t}<50\%$. Similar as the vanilla RL, the reward over the two-level framework is given as $r_{i,t}$ in \eqref{eq:reward}. Then, the objective of agent $i$ over the proposed two-level framework within $f$ time steps can be written as $R_{i}(o_{i,t},x_{i,t},o_{i,t+f}) = \mathbb{E} [\sum_{z=t}^{t+f} \gamma^{z-t} r_{i,z}]$. For each agent $i$, this process continues for $T$ time steps, emitting a new trajectory of local observations, HL actions, LL actions, and rewards: $\tau_{i} = o_{i,1},x_{i,1},a_{i,1},r_{i,1},o_{i,2},...,r_{i,T}$ over $\mathcal{O}_{i} \times \mathcal{X}_{i} \times \mathcal{A}_{i} \times \mathcal{R} \rightarrow \mathbb{R}$.

In this perspective, we consider a two-level action policy as shown in Fig. \ref{fig:h2mappo}, where agent $i$ picks up an HL action $x_{i,t}$ according to its HL policy $\mu(x|o)$, and then computes the LL actions $a_{i,t}$ according to its LL policy $pi(a|o)$ at each time step $t$ until the HL action $x_{i,t}$ switches to the other network environment. For each agent $i$, this process lasts for time steps $T$. In order to characterize the high-dimensional and continuous observation and action spaces of the HL and LL policies discussed above, an actor-critic architecture \cite{sutton2018reinforcement} is introduced for the hierarchical architecture. The actor module contains the HL policy $\mu(x|o)$ and the LL policy $\pi(a|o)$, while the critic module contains a state-value function $V(o,x)$ that specifies the expected value of selecting an HL selection $x_{i,t}$ in observation $o_{i,t}$. To deal with the complicated circumstances of the problem, deep neural networks (DNNs) are used as differentiable parameterized function approximators for both actor and critic modules. In detail, for each agent $i$, an actor network with parameters $\psi_{i}$ is constructed for training the HL policy $\mu_{\psi_{i}}(x|o)$; another actor network with parameters $\phi_{i}$ is constructed for training the LL policy $\pi_{\phi_{i}}(a|o)$; and a critic network with parameters $\theta_{i}$ is constructed for training the state-value function $V_{\theta_{i}}(o,x)$.

\subsection{Construct Hybrid Policy via MAPPO}
\label{sec:IV.B}
After selecting the HL action $x_{i,t} = \mu_{\psi_{i}}(x|o)$ for either transport routing or power scheduling, each agent $i$ at time step $t$ should execute the LL actions $a_{i,t} = \pi_{\phi_{i}}(a|o)$ to the environment. Considering that the LL routing and scheduling actions of MEG and MESS agents are in discrete and continuous spaces respectively, a hybrid policy $a_{i,t}=\{k_{i,t},c_{i,t}\} \in \mathcal{A}_{i}$ with two actor branches (networks) \cite{qiu2022hybrid} is constructed to separately compute the discrete and continuous actions in the LL for MEG and MESS agents:\footnote{It is noted here that both routing $a_{i,t}^{l,rc}$ and repairing $a_{i,t}^{r,rc}$ actions of RC agent are in discrete, we thus do not need the hybrid policy for RC agent, but adopt two categorical policies to generate $a_{i,t}^{l,rc}$ and $a_{i,t}^{r,rc}$, respectively.}
\begin{equation} \label{eq:hybrid discrete action}
    k_{i,t} = 
    \begin{cases}
    [a_{g,t}^{l,eg}] & \text{$\forall i=g \in I_{eg}$} \\
    [a_{k,t}^{l,es}] & \text{$\forall i=k \in I_{es}$} \\
    \end{cases},
\end{equation}
\begin{equation} \label{eq:hybrid continuous action}
\centering
    c_{i,t} = 
    \begin{cases}
    [a_{g,t}^{p,eg}] & \text{$\forall i=g \in I_{eg}$} \\
    [a_{k,t}^{p,es}] & \text{$\forall i=k \in I_{es}$} \\
    \end{cases}.
\end{equation}
In this case, MEG or MESS agent $i$ will choose a discrete routing action $k_{i,t}$ when the HL action $x_{i,t}$ is switched to the transport network and choose a continuous scheduling action $c_{i,t}$ when the HL action $x_{i,t}$ is switched to the power network. 

To model such action characteristics, we first generate a \texttt{softmax}$(\cdot)$ distribution for the discrete actor network parameterized by $\phi^{d}_{i}$ to output the corresponding probabilities for all potential routing behaviors, this categorical policy $\texttt{softmax}(o)=k_{i,t}=\pi_{\phi^{d}_{i}}(k|o)$ is then sampled for the optimal action $k_{i,t}$ in observation $o_{i,t}$. The softmax activation function takes in the local observation $o$ and returns the probability scores of all possible discrete routing actions. In general, the equation of \texttt{softmax}$(\cdot)$ distribution is given as
\begin{equation} \label{eq:softmax}
    \texttt{softmax}(o)_{d} = \frac{e^{o_{d}}}{\sum_{j=1}^{K}e^{o_{j}}}, \forall d \in 1,...,K,
\end{equation}
where $K$ indicates the number of dimensions of LL discrete routing action $k_{i,t}$ in \eqref{eq:hybrid discrete action}. In principle, function \eqref{eq:softmax} applies the standard exponential function to each element $o_{d}$ of the input local observation $o$ and normalizes these values by dividing by the sum of all these exponentials; this normalization ensures that the sum of all elements' output probabilities is 1.

We then generate a Gaussian distribution $f(\cdot)$ for the continuous actor network parameterized by $\phi^{c}_{i}$ to output the corresponding mean and variance for all scheduling behaviors, the stochastic policy $f(o) = c_{i,t} = \pi_{\phi^{c}_{i}}(c|o)$ is then sampled for the optimal action $c_{i,t}$ in observation $o_{i,t}$. A Gaussian distribution is a type of continuous probability distribution for a real-valued random variable. The general form of its probability density function is
\begin{equation} \label{eq:gaussian}
    f(o) = \frac{1}{\sigma {\sqrt{2\pi}}} e^{-{\frac{1}{2}} \left({\frac{o-\mu}{\sigma}}\right)^{2}}
\end{equation}
where the parameter $\mu$ is the mean of the distribution, while the parameter $\sigma$ is its standard deviation. The objective of the continuous actor network parameterized by $\phi^{c}_{i}$ is learning the parameters $\mu$ and $\sigma$ in \eqref{eq:gaussian} given input local observation $o$.

The discrete policy $\pi_{\phi^{d}_{i}}$ and continuous policy $\pi_{\phi^{c}_{i}}$ are then updated independently using the MAPPO algorithm \cite{yu2021surprising}, which minimizes their clipped surrogate objective to restrict the policy update:
\begin{equation} \label{eq:hybrid ppo discrete clip policy}
    L_{i,t}^{\text{CLIP}}(\phi^{d}_{i}) = \hat{\mathbb{E}}_{t} \big[\min(\zeta_{i,t}^{d}\hat{A}_{i,t}, \text{clip}(\zeta_{i,t}^{d},1-\epsilon,1+\epsilon)\hat{A}_{i,t} ) \big],
\end{equation}
\begin{equation} \label{eq:hybrid ppo continuous clip policy}
    L_{i,t}^{\text{CLIP}}(\phi^{c}_{i}) = \hat{\mathbb{E}}_{t} \big[\min(\zeta_{i,t}^{c}\hat{A}_{i,t}, \text{clip}(\zeta_{i,t}^{c},1-\epsilon,1+\epsilon)\hat{A}_{i,t} ) \big],
\end{equation}
where the former term in the operator $\min\{\cdot\}$ indicates the normal policy gradient while the later term in the operator $\min\{\cdot\}$ trims the policy gradient by clipping the probability ratio $\zeta_{i,t}^{d},\zeta_{i,t}^{c}$ between $[1-\epsilon,1+\epsilon]$. The hyperparameter $\epsilon \in [0,1]$ is used to truncate the gradient update of the new policy from the old version. In other words, the advantage function $\hat{A}_{i,t}$ will be clipped if the probability ratio between the new and old policies goes beyond the range $[1-epsilon,1+epsilon]$. In hybrid policy, the probability ratio $\zeta_{i,t}^{d}$ only takes into account the discrete policy, whereas the probability ratio $\zeta_{i,t}^{c}$ only takes into account the continuous policy. Specifically,
\begin{equation} \label{eq:hybrid ppo two policies}
    \zeta_{i,t}^{d} = \frac{\pi_{\phi^{d}_{i}}(k_{i,t}|o_{i,t})}{\pi_{\phi^{d}_{i}\text{old}}(k_{i,t}|o_{i,t})} \quad \!\!\text{and}\!\! \quad \zeta_{i,t}^{c} = \frac{\pi_{\phi^{c}_{i}}(c_{i,t}|o_{i,t})}{\pi_{\phi^{c}_{i}\text{old}}(c_{i,t}|o_{i,t})}.
\end{equation}

Then, the HL discrete policy $\mu_{\psi_{i}}(x|o)$ can be updated similarly as the LL discrete policy $\pi_{\phi^{d}_{i}}(k|o)$ in \eqref{eq:hybrid ppo discrete clip policy}, while the probability ratio of HL policy $\zeta_{i,t}^{x}$ can also be derived similarly as the LL discrete one $\zeta_{i,t}^{d}$ in \eqref{eq:hybrid ppo two policies}. 

In addition, the generalized advantage function $\hat{A}_{i,t}$ can be expressed as
\begin{equation} \label{eq:hybrid ppo advantage}
    \hat{A}_{i,t} = \delta_{i,t} + \gamma \delta_{i,t+1} + \dots + \gamma^{T-t+1} \delta_{i,T-1},
\end{equation}
\begin{equation} \label{eq:hybrid ppo value function}
    \delta_{i,t} = r_{i,t} + \gamma V_{\theta_{i}}(o_{1:I,t+1},x_{1:I,t+1}) - V_{\theta_{i}}(o_{1:I,t},x_{1:I,t}),
\end{equation}
where $V_{\theta_{i}}(o,x)$ is the state-value function incorporating centralized training with the local observations $o_{1:I}$ and HL actions $x_{1:I}$ of all agents, which is approximated by a critic network parameterized by $\theta_{i}$ introduced in Section \ref{sec:IV.A}. It should be mentioned that providing all local information to the critic network can stabilize learning and foster coordinated behaviors for all local agents \cite{yu2021surprising}.

\subsection{Abstract System Dynamics}
\label{sec:IV.C}
However, the centralized critic network taking all agents' local information may raise problems. First, the shared information among all agents can destroy their privacy, since these decentralized MPSs and RCs are not willing to exchange their dispatch behaviors with each other \cite{lei2018routing}. Second, a POMDP may not be reduced to an MDP by concatenating all local information in centralised training since there may be crucial information (e.g., load shedding quantity) that is not noticed by any of the agents during training. This paper thus abstracts the system's global dynamics (e.g., load shedding quantity) via an embedded function $\xi_{i}$ and approximates a new multi-agent joint state-value function as
\begin{equation} \label{eq:h2mappo index}
    V_{\theta_{i}}(o_{1:I},x_{1:I}) = V_{\theta_{i}}(o_{i},x_{i},\xi_{i}),
\end{equation}
which inputs individual agent's local observation $o_{i}$, HL action $x_{i}$, and embedded function $\xi_{i}$. Specifically, $\xi_{i}$ indicates how much each agent $i$ contributes to the system overall load restoration, which can be expressed as
\begin{equation} \label{eq:index contribution} 
    \xi_{i} =
    \begin{cases}
    |P_{j}^{rc}| / \sum_{d \in D}(P^{ed}_{d}-P^{ls}_{d}) & \text{$\forall i = j \in I_{rc}$} \\
    P_{g}^{eg} / \sum_{d \in D}(P^{ed}_{d}-P^{ls}_{d}) & \text{$\forall i = g \in I_{eg}$} \\
    |P_{k}^{esd}| / \sum_{d \in D}(P^{ed}_{d}-P^{ls}_{d}) & \text{$\forall i = k \in I_{es}$}
    \end{cases},
\end{equation}
where $P_{i,t}^{rc}$ represents the power flow through the repaired line.

In this setting, $\xi_{i}$ can be assumed to abstract the local observations of all other agents, e.g., $P_{i',t}^{ed}, P_{i',t}^{pv}, S_{i',t}^{ln}, \forall i' \in I(i')$, where $I(i')$ indicates the set of all other agents $i'$ apart from $i$. Furthermore, $\xi_{i}$ can reflect the status of agents supporting system resilience, i.e., the higher value of $\xi_{i}$ denote the better performance of load restoration, and vice versa. As a consequence, each agent can make informative actions based on the abstracted knowledge $\xi_{i}$ of local observations and actions from all other agents, albeit not directly obtaining their local information and dispatch behaviors, therefore safeguarding the privacy and enhancing scalability.\footnote{It should be noted that directly integrating function $\xi_{i}$ into the current observation $o_{i,t}$ is unimplementable, since they can be only accessible once MGCC has solved the AC-OPF algorithm, which requires all agents' actions conditioned on the current local observations. However, the function $\xi_{i}$ can be used in the critic training process, which is carried out after integrating them into the state-value function in \eqref{eq:h2mappo index}.}

\subsection{Training Process}
\label{sec:IV.D}
During the training process, H2MAPPO runs for all agents by their individual HL and LL policies $\mu_{\psi_{i}}(x|o),\pi_{\phi_{i}}(a|o)$ through $T$ time steps, while collecting the trajectories $\tau_{i}$ (including the function $\xi_{i}$) from the interactions with the environment. Then, the agents can use the gathered trajectories to calculate the discounted reward-to-go $\hat{R}_{\iota,t} = \sum_{h=t}^{T} \gamma^{h-t} r_{\iota,h}$ and the advantage function $\hat{A}_{\iota,t}$ according to the abstracted state-value function $V_{\theta_{i}}(o_{\iota,t},x_{\iota,t},\xi_{\iota,t})$ for each trajectory $\iota$ and time step $t$, where a batch of trajectories are taken from the buffer $\mathcal{J} = \{\tau_{\iota}\} \sim \mathcal{F}$. Then, three actor networks are trained by maximising their objectives as follows:
\begin{equation} \label{eq:hl train}
    \mathcal{L}(\psi_{i}) \!=\! \frac{1}{J} \! \sum_{\iota=1}^{J} \! \min \big( \zeta_{\iota,t}^{x} \hat{A}_{\iota,t}, \text{clip}(\zeta_{\iota,t}^{x},\!1-\epsilon,\!1+\epsilon)\hat{A}_{\iota,t} \big),
\end{equation}
\begin{equation} \label{eq:ll discrete train}
    \mathcal{L}(\phi^{d}_{i}) \!=\! \frac{1}{J} \! \sum_{\iota=1}^{J} \! \min \big( \zeta_{\iota,t}^{d} \hat{A}_{\iota,t}, \text{clip}(\zeta_{\iota,t}^{d},\!1-\epsilon,\!1+\epsilon)\hat{A}_{\iota,t} \big),
\end{equation}
\begin{equation} \label{eq:ll continuous train}
    \mathcal{L}(\phi^{c}_{i}) \!=\! \frac{1}{J} \! \sum_{\iota=1}^{J} \! \min \big( \zeta_{\iota,t}^{c} \hat{A}_{\iota,t}, \text{clip}(\zeta_{\iota,t}^{c},\!1-\epsilon,\!1+\epsilon)\hat{A}_{\iota,t} \big),
\end{equation}
where $J$ indicates the training batch size. The critic network is trained with the objective of minimizing the mean-squared error loss function:
\begin{equation} \label{eq:critic train}
    \mathcal{L}(\theta_{i}) = \frac{1}{J} \sum_{\iota=1}^{J} \min \big( V_{\theta_{i}}(o_{\iota,t},x_{\iota,t},\xi_{\iota,t}) - \hat{R}_{\iota,t} \big)^{2}.
\end{equation}

Given the above optimizations, the network weights of three actors and one critic can be updated as below:
\begin{equation} \label{eq:hl weights}
    \psi_{i} \leftarrow \psi_{i} + \alpha^{\psi} \nabla_{\psi_{i}}\mathcal{L}(\psi_{i}), 
\end{equation}
\begin{equation} \label{eq:ll discrete weight}
    \phi^{d}_{i} \leftarrow \phi^{d}_{i} + \alpha^{\phi^{d}} \nabla_{\phi^{d}_{i}}\mathcal{L}(\phi^{d}_{i}), 
\end{equation}
\begin{equation} \label{eq:ll discrete weight}
    \phi^{c}_{i} \leftarrow \phi^{c}_{i} + \alpha^{\phi^{c}} \nabla_{\phi^{c}_{i}}\mathcal{L}(\phi^{c}_{i}), 
\end{equation}
\begin{equation} \label{eq:critic weight}
    \theta_{i} \leftarrow \theta_{i} + \alpha^{\theta} \nabla_{\theta_{i}}\mathcal{L}(\theta_{i}),
\end{equation}
where $\alpha^{\psi},\alpha^{\phi^{d}},\alpha^{\phi^{d}},\alpha^{\theta}$ indicate the learning rates of the gradient ascent/descent algorithms for actor/critic networks. Finally, the pseudo-code of H2MAPPO is presented as below:

\begin{algorithm}
\setstretch{1.00}
\algsetup{linenosize=\small}
\small
\caption{Training process of H2MAPPO for $I$ agents}
\label{algorithm:h2mappo}
\begin{algorithmic}[1]
\STATE Initialize weights $\psi_{i},\phi^{d}_{i},\phi^{c}_{i},\theta_{i}$ for actor and critic networks
\STATE Set learning rates $\alpha^{\psi},\alpha^{\phi^{d}},\alpha^{\phi^{c}},\alpha^{\theta}$
\FOR{episode (i.e., day) $epi = 1$ to $E$}
\STATE Initialize the global state $s_{0}$ and local observation $o_{i,0}$
\STATE For each agent $i$, sets an empty buffer $\mathcal{F}=\{\}$
\STATE For each agent $i$, sets an empty trajectory $\tau_{i}=[]$
\STATE For each agent $i$, selects HL action $x_{i,0}$ in observing $o_{i,0}$
\FOR{time step (i.e., 1 hour) $t = 1$ to $T$}
\REPEAT
\FOR{agent (i.e., RC, MEG, MESS) $i = 1$ to $I$}
\STATE Selects LL discrete action $a_{i,t}=k_{i,t}$ (if HL action is for transport) or continuous action $a_{i,t}=c_{i,t}$ (if HL action is for power)
\ENDFOR
\STATE Execute all agents' actions $a_{1:I,t}$ to the environment, including both transport and power networks
\STATE MGCC runs the AC-OPF algorithm once collecting all agents' dispatches (RCs' repairing decision, MEGs' power generation, and MESSs' charging and discharging power) and calculates the embedded function $\xi_{i,t}$ (if HL action is for power)
\FOR{agent (i.e., RC, MEG, MESS) $i = 1$ to $I$}
\STATE Observes reward $r_{i,t}$ and next observation $o_{i,t+1}$ 
\STATE Stores one sample experience to trajectory $\tau_{i} \mathrel{+}= [o_{i,t},x_{i,t},a_{i,t},r_{i,t},\xi_{i,t}]$
\WHILE{time step $t \,\%\, J = 0$}
\STATE Collects a set of trajectories $\tau_{\iota}$ from buffer $\mathcal{F}$, then computes advantage function $\hat{A}_{\iota,t}$ and discounted reward-to-go $\hat{R}_{\iota,t}$
\STATE Updates network weights $\psi,\phi^{d},\phi^{c},\theta$ in \eqref{eq:hl weights}-\eqref{eq:critic weight}
\ENDWHILE
\ENDFOR
\STATE Update local observation $o_{i,t} \leftarrow o_{i,t+1}$
\UNTIL HL action $x_{i,t}$ is switched in new observation $o_{i,t}$
\STATE Update HL action $x_{i,t} \leftarrow x_{i,t+1}$
\ENDFOR
\ENDFOR
\end{algorithmic}
\end{algorithm}
\setlength{\textfloatsep}{5pt}

\subsection{Test Process}
\label{sec:IV.E}
The training process lasts for $E$ episodes until the trained H2MAPPO method is being converged. Once in the test process, we firstly collect the weight parameters $\psi_{i}$ of HL policy network as well as $\phi^{d}_{i},\phi^{c}_{i}$ of LL discrete and continuous policy networks respectively trained in Algorithm \ref{algorithm:h2mappo}. The critic network is no longer required in the test process. For each time step in the test days $F$, each (RC, MEG, MESS) agent $i$ observes its current local observation $o_{i,t}$ and accordingly executes the HL action $x_{i,t}$ for environment switch as well as the LL discrete or continuous action $a_{i,t}=\{k_{i,t},c_{i,t}\}$ for behaving in transport or power network. Those decisions are then mapped to the operation models of the coupled power-transport network (environment), transiting to the next state and time step. Each agent can also obtain its individual reward from the environment. Overall, the test process of the proposed H2MAPPO method is presented as below:

\begin{algorithm}
\setstretch{1.00}
\algsetup{linenosize=\small}
\small
\caption{Test process of H2MAPPO for $I$ agents}
\label{algorithm:test}
\begin{algorithmic}[1]
\STATE Load the weights $\psi_{i},\phi^{d}_{i},\phi^{c}_{i}$ trained by Algorithm \ref{algorithm:h2mappo}
\FOR{test day = $1:F$}
\STATE Initialize the global state $s_{0}$ and local observation $o_{i,0}$
\STATE For each agent $i$, selects HL action $x_{i,0}$ in observing $o_{i,0}$
\FOR{time step (i.e., 1 hour) $t = 1$ to $T$}
\REPEAT
\FOR{agent (i.e., RC, MEG, MESS) $i = 1$ to $I$}
\STATE Selects LL discrete action $a_{i,t}=k_{i,t}$ (if HL action is for transport) or continuous action $a_{i,t}=c_{i,t}$ (if HL action is for power)
\ENDFOR
\STATE Execute all agents' actions $a_{1:I,t}$ to the environment, including both transport and power networks
\STATE MGCC runs the AC-OPF algorithm once collecting all agents' dispatches (RCs' repairing decision, MEGs' power generation, and MESSs' charging and discharging power) and calculates the load restoration $P_{d,t}^{ed}$
\FOR{agent (i.e., RC, MEG, MESS) $i = 1$ to $I$}
\STATE Observes reward $r_{i,t}$ and next observation $o_{i,t+1}$ 
\ENDFOR
\STATE Update local observation $o_{i,t} \leftarrow o_{i,t+1}$
\UNTIL HL action $x_{i,t}$ is switched in new observation $o_{i,t}$
\STATE Update HL action $x_{i,t} \leftarrow x_{i,t+1}$
\ENDFOR
\ENDFOR
\end{algorithmic}
\end{algorithm}
\setlength{\textfloatsep}{5pt}

\section{Case Studies}
\label{sec:V}
\subsection{Experimental Setup}
\label{sec:V.A}
\subsubsection{Network setup}
\begin{figure}[t!]
\centering
\subfigure[The modified IEEE 33-bus power distribution network]{\includegraphics[width=0.485\textwidth]{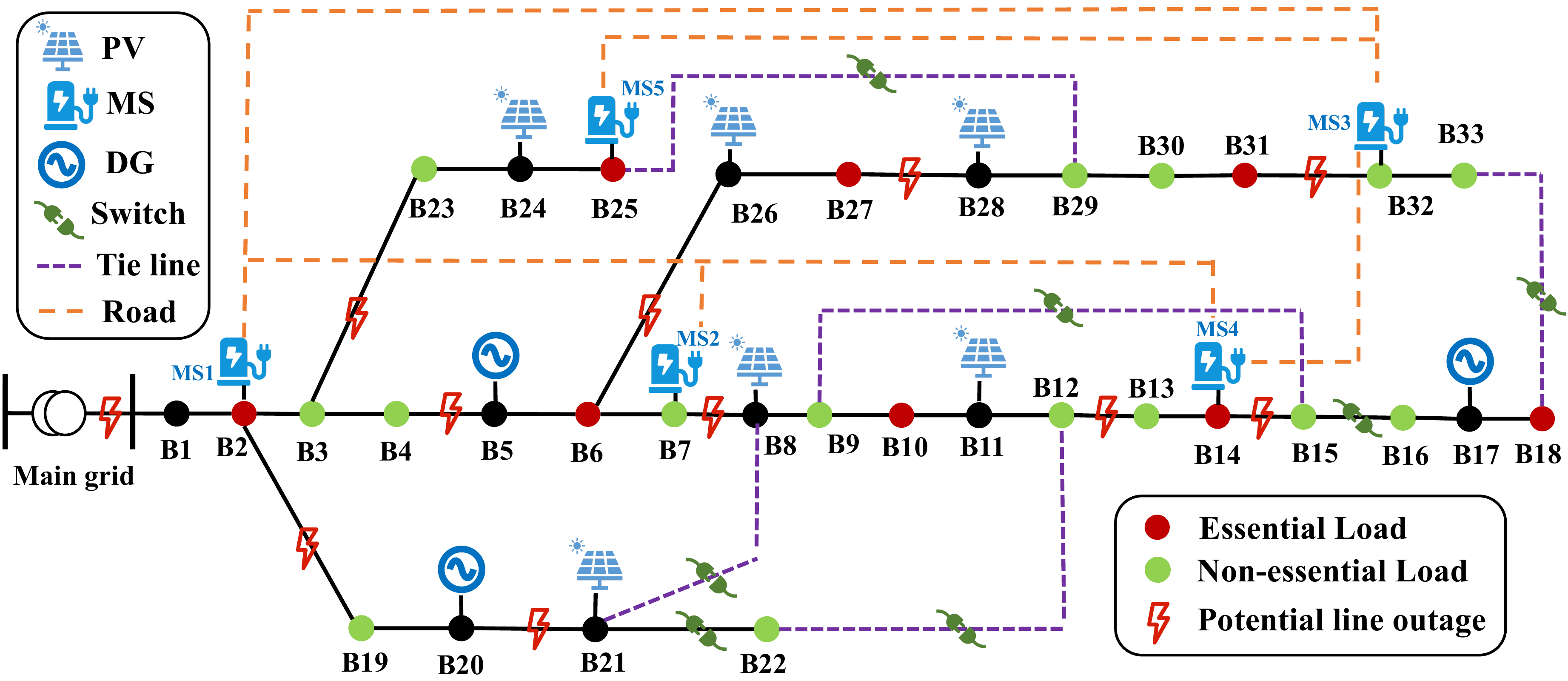}}
\subfigure[MPSs' transport network]{\includegraphics[width=0.185\textwidth]{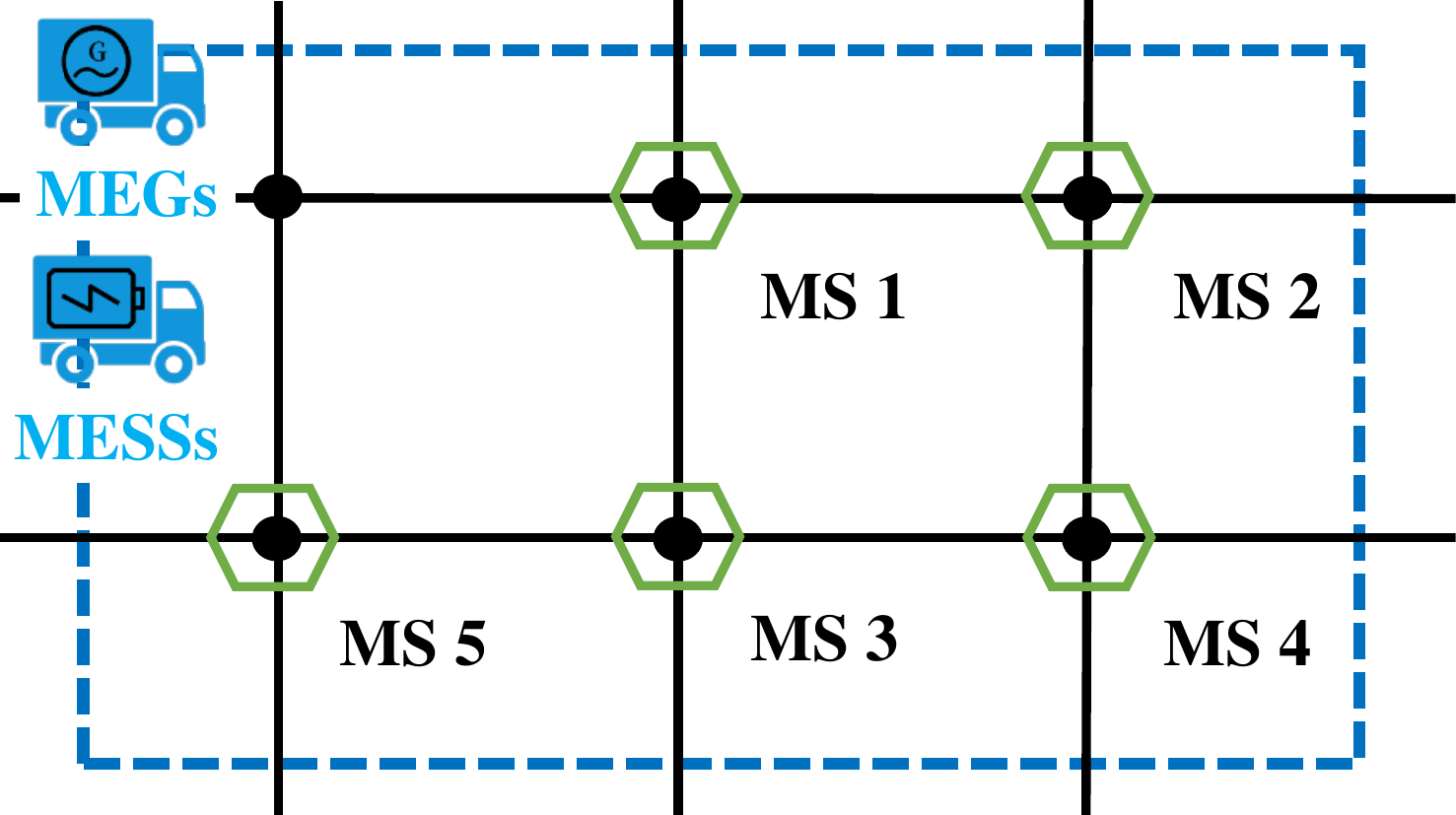}}
\subfigure[RCs' transport network]{\includegraphics[width=0.275\textwidth]{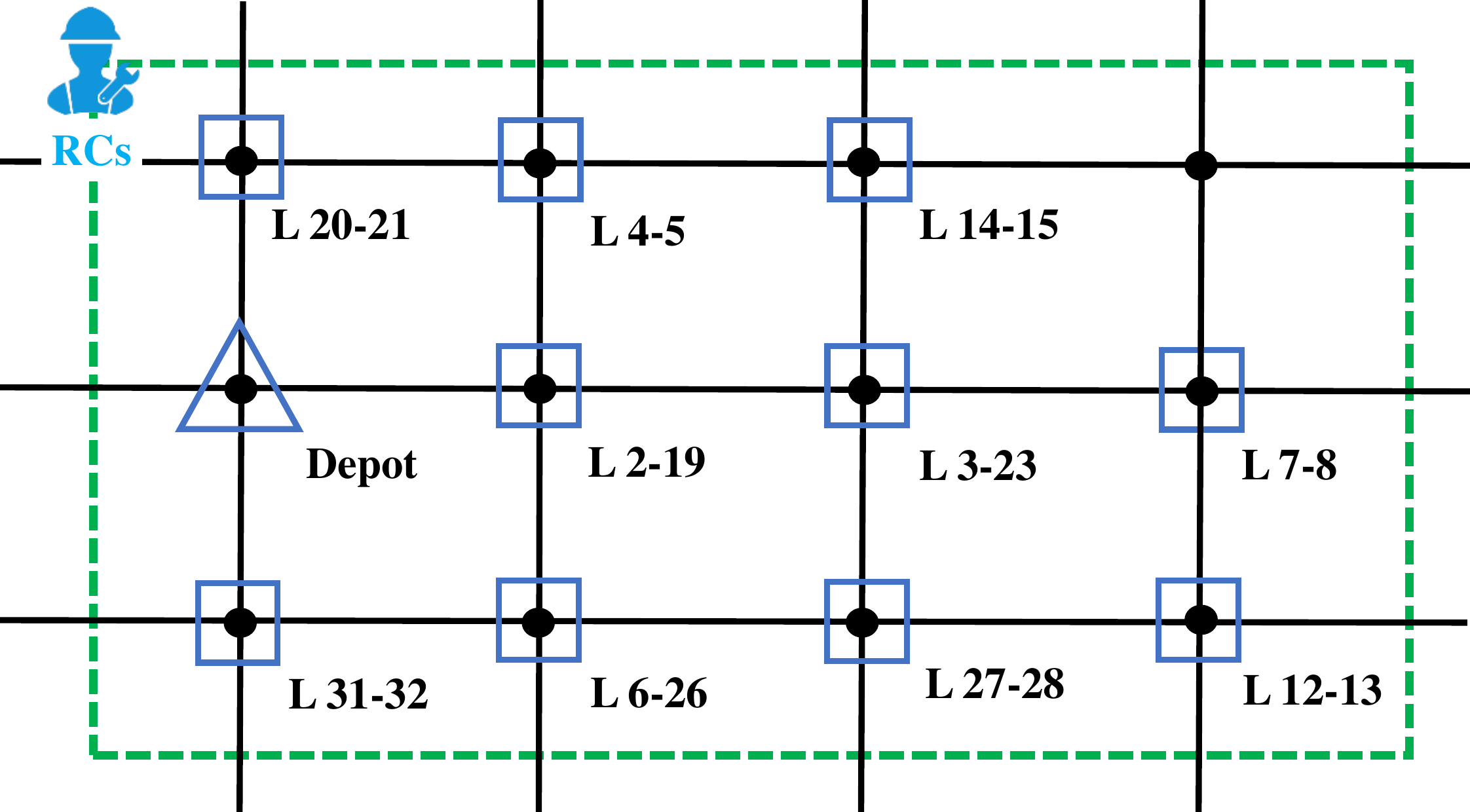}}
\caption{The coupled power-transport network utilized for case studies: (a) the modified 33-bus power distribution network, (b) the transport network with MSs for MPSs, (c) the transport network with damaged components for RCs.}
\label{fig:33bus}
\end{figure}

To assess the effectiveness of the proposed MARL method in capturing realistic MPS and RC dispatch behaviors, a modified IEEE 33-bus power network, as shown in Fig. \ref{fig:33bus}. The power network has 8 essential loads, 15 non-essential loads, 6 PVs, 3 DGs, and 5 MSs. Mobile resources deployed for load restoration include 1 MEG, 1 MESS and 1 RC. In the transport network, we assume that these resources can move to any candidate node through their routing characteristics, where detailed transport network structures between these candidate nodes (i.e., MSs for MPSs and damaged components for RCs) can be found in Fig. \ref{fig:33bus}.

To capture the impact of extreme events, we assume that multiple line outages can happen in the power network, as depicted in Fig. \ref{fig:33bus}. Specifically, the Monte Carlo sampling technique can be used to generate a manageable number of scenarios based on the fragility curve suggested in \cite{7559799}. For each episode, a random outage scenario will be sampled from the fragility curve to represent the damaged conditions. Within the distribution network as shown in Fig. \ref{fig:33bus}, the potential lines with higher damaged probabilities are easier to be selected into the outage scenario.

\begin{table}[t!]
\centering
\footnotesize
\setlength{\abovecaptionskip}{10pt}
\renewcommand\arraystretch{1.05}
\caption{Technical Parameters of 3 DGs}
\setlength{\tabcolsep}{1.98mm}{
\begin{tabular}{ | c | c | c | c | c | }
	  \hline
	  DG & \begin{tabular}[c]{@{}c@{}}$\underline{P}^{dg}$ (kW)\end{tabular} & \begin{tabular}[c]{@{}c@{}}$\overline{P}^{dg}$ (kW)\end{tabular} & \begin{tabular}[c]{@{}c@{}}$\underline{Q}^{dg}$ (kVAR)\end{tabular} & \begin{tabular}[c]{@{}c@{}}$\overline{Q}^{dg}$ (kVAR)\end{tabular} \\ \hline
      1 & 0 & 200 & -67 & 100 \\ \hline
      2 & 0 & 300 & -100 & 150 \\ \hline
      3 & 0 & 400 & -133 & 200 \\ \hline
\end{tabular}}
\label{table:para_dg}
\vspace{0.5em}
\end{table}

\begin{table}[t!]
\footnotesize
\centering
\setlength{\abovecaptionskip}{10pt}
\renewcommand\arraystretch{1.15}
\caption{Technical Parameters of MESS, MEG and RC}
\setlength{\tabcolsep}{2.58mm}{
\begin{tabular}{ |c|c|c|c|c|c| }
    \hline
    \multicolumn{2}{|c|}{RC} & \multicolumn{2}{c|}{MEG} & \multicolumn{2}{|c|}{MESS} \\ 
    \hline
    $\overline{S}^{rc}$ (unit) & 10 & $\overline{P}^{eg}$ (kW) & 150 & $\overline{P}^{es}$ (kW) & 100 \\
    \hline
    $RS^{rc}$ (unit) & [2,3] & $\overline{Q}^{eg}$(kVAR) & 75 & $\overline{E}^{es}$ (kWh) & 400 \\
    \hline
    $RT^{rc}$ (h) & [1,4] & $\underline{Q}^{eg}$(kVAR) & -50 & $\eta^c/\eta^d$ (\%) & 90 \\
    \hline
\end{tabular}}
\label{table:para_mps}
\vspace{0.5em}
\end{table}

\subsubsection{Data Descriptions}
\label{sec:V.A.2}
Case studies are conducted based on a real-world dataset from the Ausgrid \cite{ratnam2017residential}. The one-year PV generation and residential load data are collected, and then split into train (11 months) and test (1 month) sets for MARL method. To reflect the load distinction, 30\% of loads are assumed to be essential featuring a high shedding cost at 2.5 \pounds/kW, while the other 70\% are non-essential and have a shedding cost at 1.5 \pounds/kW. Tables \ref{table:para_dg} and \ref{table:para_mps} present the technical parameters of 3 static DGs and 3 mobile resources, respectively.

\subsubsection{Benchmarks}
\label{sec:V.A.3}
The proposed H2MAPPO is compared with four benchmarks, including two MARL methods and two optimization methods: i) \textbf{IPPO}: each agent independently employs PPO , introducing a Gaussian policy with two dimensions, i.e., continuous scheduling action and discrete routing action by separating the continuous space into a finite set of segments; ii) \textbf{MAPPO}: based on IPPO, all local observations $o_{1:I}$ are concatenated by each agent to formulate its individual value function $V_{i}(o_{1:I})$; iii) \textbf{MPC}: MGCC employs a stochastic model predictive control (MPC) approach to solve a rolling optimization problem, which is constructed with the objective function \eqref{eq:obj} and constraints \eqref{eq:ev flow1}-\eqref{eq:rc_5} and \eqref{eq:api}-\eqref{eq:radial_4}, assuming the perfect information of the power-transport network, mobile resource models, and all technical parameters; iv) \textbf{MILP}: MGCC employs a central deterministic mixed-integer linear programming (MILP) for the daily optimization problem with the objective function \eqref{eq:obj} and constraints \eqref{eq:ev flow1}-\eqref{eq:rc_5} and \eqref{eq:api}-\eqref{eq:radial_4}, assuming the perfect knowledge of the system uncertainties. 

\subsubsection{Implementations}
\label{sec:V.A.4}
The training process of both actor and critic networks use the Adam optimizer with learning rates $\alpha^{\psi}=\alpha^{\phi^{d}}=\alpha^{\phi^{c}}=10^{-4}$ and $\alpha^{\theta}=10^{-3}$. The batch size $J=24$ corresponds to 24 environment steps per episode. The discount rate $\gamma=0.99$ and the clip rate $\epsilon=0.2$. Multi-layer Perceptrons (MLPs) constructed by two hidden layers (128 and 64 units) with \texttt{ReLU} activation function are utilized for all networks. \texttt{Softmax} activation function is utilized to construct the categorical policy for both HL and LL discrete (routing and RC repair) actors. \texttt{Tanh} and \texttt{Softplus} activation functions are utilized to respectively construct a Gaussian policy with mean and standard deviation for the LL continuous (MPS scheduling) actor. Overall, the detailed specifications of actor and critic networks for three utilized MARL methods are presented in Table \ref{table:marl structure}. Both IPPO and MAPPO feature a single actor-critic network, the difference is that the input of critic network for IPPO is the individual local observation while the input of critic network for MAPPO is the all agents' local observations. For our proposed H2MAPPO, the network structure becomes more complicated. Specifically, there are four separate actor networks, indicating the HL actor for switch option; the first LL actor for routing decisions of both MPSs and RCs; the second LL actor for scheduling decisions of MPSs; and the third LL actor for repairing decisions of RCs. For all MARL methods, we run 5,000 episodes with the same 10 random seeds.

\begin{table}[h!]
\centering
\footnotesize
\setlength{\abovecaptionskip}{12pt}
\renewcommand\arraystretch{1.05}
\caption{General Specifications of Three MARL Methods.}
\scalebox{0.692}{
\begin{tabular}{ |l|c|c| }
    \hline
    Mechanism & Network & Structure \\
    \hline
    \multirow{2}{*}{IPPO} 
        & Actor & \texttt{linear(o\_dim, 128) → ReLU() x2 → tanh+softplus(64, 2)} \\
        & Critic & \texttt{linear(o\_dim, 128) → ReLU() x2 → linear(64, 1)} \\
    \hline
    \multirow{2}{*}{MAPPO} 
        & Actor & \texttt{linear(o\_dim, 128) → ReLU() x2 → tanh+softplus(64, 2)} \\
        & Critic & \texttt{linear(o\_dim × |I|, 128) → ReLU() x2 → linear(64, 1)} \\
    \hline
    \multirow{5}{*}{H2MAPPO} 
        & HL actor & \texttt{linear(o\_dim, 128) → ReLU() x2 → softmax(64, 2)} \\
        & LL actor (route) & \texttt{linear(o\_dim, 128) → ReLU() x2 → softmax(64, 4)} \\
        & LL actor (MPS) & \texttt{linear(o\_dim, 128) → ReLU() x2 → tanh+softplus(64, 1)} \\
        & LL actor (RC) & \texttt{linear(o\_dim, 128) → ReLU() x2 → softmax(64, 2)} \\
        & Critic & \texttt{linear(o\_dim+2, 128) → ReLU() x2 → linear(64, 1)} \\
    \hline
\end{tabular}}
\label{table:marl structure}
\vspace{0.5em}
\end{table}

\subsection{Performance Evaluation}
\label{sec:V.B}
\begin{figure}[!h]
\centering
\includegraphics[width=0.385\textwidth]{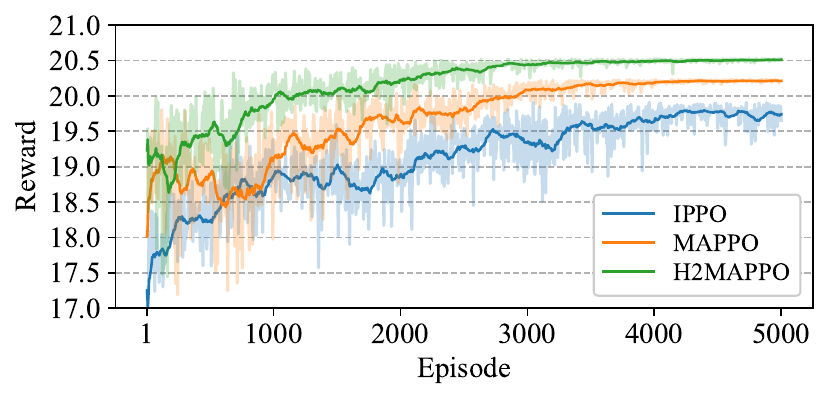}
\vspace{-0.75em}
\caption{Episodic reward over 5,000 episodes for different MARL methods.}
\label{fig:train reward} 
\end{figure}

In this subsection, the training and test performance of three investigated MARL methods are evaluated. Fig. \ref{fig:train reward} depicts the evolution of episodic reward over 5,000 training episodes, where the solid lines and the shaded areas respectively depict the moving average over 100 episodes and the oscillations of the original reward. Furthermore, their corresponding averaged episodic training time as well as the averaged number of episodes and averaged total training time required to reach convergence are collected in Table \ref{table:train result}. Finally, we also collect the averaged resilience index ($\lambda$) and computation time over the 31 test days for three MARL and two optimization methods in Table \ref{table:test result}.

\begin{table}[h!]
\footnotesize
\centering
\setlength{\abovecaptionskip}{10pt}
\renewcommand\arraystretch{1.00}
\caption{Computational Performance for Different MARL Methods}
\begin{threeparttable}[t]
\setlength{\tabcolsep}{2.98mm}{
\begin{tabular}{ |l|c|c|c| }
 	  \hline
      Method & IPPO & MAPPO & H2MAPPO \\
      \hline
      Episodic training time (s.) & 1.83 & 2.14 & 2.65 \\
      \hline
      Number of episodes (\#) & 5,000\tnote{*} & 4,000 & 3,200 \\
      \hline
      Total training time (hr.) & 2.54\tnote{*} & 2.38 & 2.36 \\
      \hline
\end{tabular}}
\begin{tablenotes}
    \item[*] Fail to reach convergence within 5,000 episodes.
\end{tablenotes}
\end{threeparttable}
\label{table:train result}
\end{table}

\begin{table}[!h]
\footnotesize
\centering
\setlength{\abovecaptionskip}{10pt}
\renewcommand\arraystretch{1.00}
\caption{Averaged Resilience Index and Computation Time over 31 Test Days for Different MARL and Optimization Methods}
\setlength{\tabcolsep}{1.68mm}{
\begin{tabular}{ |l|c|c|c|c|c| }
 	  \hline
      Method & IPPO & MAPPO & H2MAPPO & MPC & MILP \\
      \hline
      Index-$\lambda$ & 19.56 & 20.15 & 21.34 & 19.93 & 22.12 \\
      \hline
      Computation (sec.) & 0.59 & 0.49 & 0.54 & 1028.93 & 76.21 \\
      \hline
\end{tabular}}
\label{table:test result}
\end{table}

The first observation we notice from Fig. \ref{fig:train reward} is that IPPO (blue) has the most unstable and oscillatory training performance, thereby obtaining the lowest reward level and failing to reach optimum. The independent learning method of IPPO, which concentrates on local information while disregarding the other agents, is considered to be the major cause of this instability issue, making the environment non-stationary. As a consequence, by concatenating all agents' information, MAPPO (orange) can effectively mitigate such non-stationarity and thus display better stability performance. However, MAPPO has inadequate policy quality due to the simple division of action space, which dramatically reduces its efficacy in handling scheduling actions in the power network, resulting in sub-optimum. Additionally, in the absence of a hierarchical architecture, each resource agent acquires routing and scheduling actions simultaneously, which may result in ineffective critic network learning, because there is always one meaningless action in the environment. In this case, the proposed H2MAPPO (green) can address the above issues by 1) utilizing embedded function $\xi$ to learn system dynamics; 2) employing the hybrid policy to generate both continuous and discrete actions separately; and 3) introducing a hierarchical architecture that allows agents to adaptively switch to suitable environment status (power network or transport network). 

We further assess the computational performance of three MARL methods during the training process. Table \ref{table:train result} shows that IPPO has the shortest episodic training time (since it only requires training one actor network to compute both routing and scheduling/repairing actions, eliminating the need for hierarchical architecture), followed by MAPPO (since it takes all agents' local observations as the inputs of the centralized critic network), and H2MAPPO (since it needs to train four actor networks rather than the single actor network in IPPO and MAPPO). Additionally, we see that H2MAPPO (around 3,200 episodes) demonstrates a faster convergence rate than MAPPO (around 4,000 episodes). This is because the critic network incorporates an embedded function $\xi_{i}$ that can stabilize the training performance and obtain a faster learning algorithm. Due to its instability issue, IPPO fails to reach convergence within 5,000 episodes. Finally, our proposed H2MAPPO (2.36 hrs) costs the similar computational time to MAPPO (2.38 hrs) but obtains a better policy quality (i.e., higher resilience level).

Regarding test performance in Table \ref{table:test result}, the proposed H2MAPPO achieves a near-to optimal performance (3.53\% lower than MILP), and outperforms IPPO, MAPPO, and MPC in terms of the averaged resilience index over 31 test days by 9.10\%, 5.91\%, and 7.07\%, respectively. On the other hand, all three MARL methods can be deployed in real-time around 0.5 sec., while the optimization-based MPC and MILP requires around 1000 sec. and 75 sec. averaged per day. It is worth noting that real-time control is important to the resilient MG operation problem due to the demand of a fast response time. 

\subsection{Analysis of Dispatch Behaviors and Switch Operations}
\label{sec:V.C}
After evaluating the MARL performance, this section validates the learned policy of H2MAPPO for dispatch behaviors of three resources, while the MG switch operations and load conditions are also involved. A scenario with 6 line outages (lines $4-5$, $14-15$, $2-19$, $3-23$, $6-26$ and $31-32$) is selected here. Additionally, serious traffic congestion mainly happens in the afternoon during the rush hours. 

\begin{figure}[h!]
\centering
\includegraphics[width=0.475\textwidth]{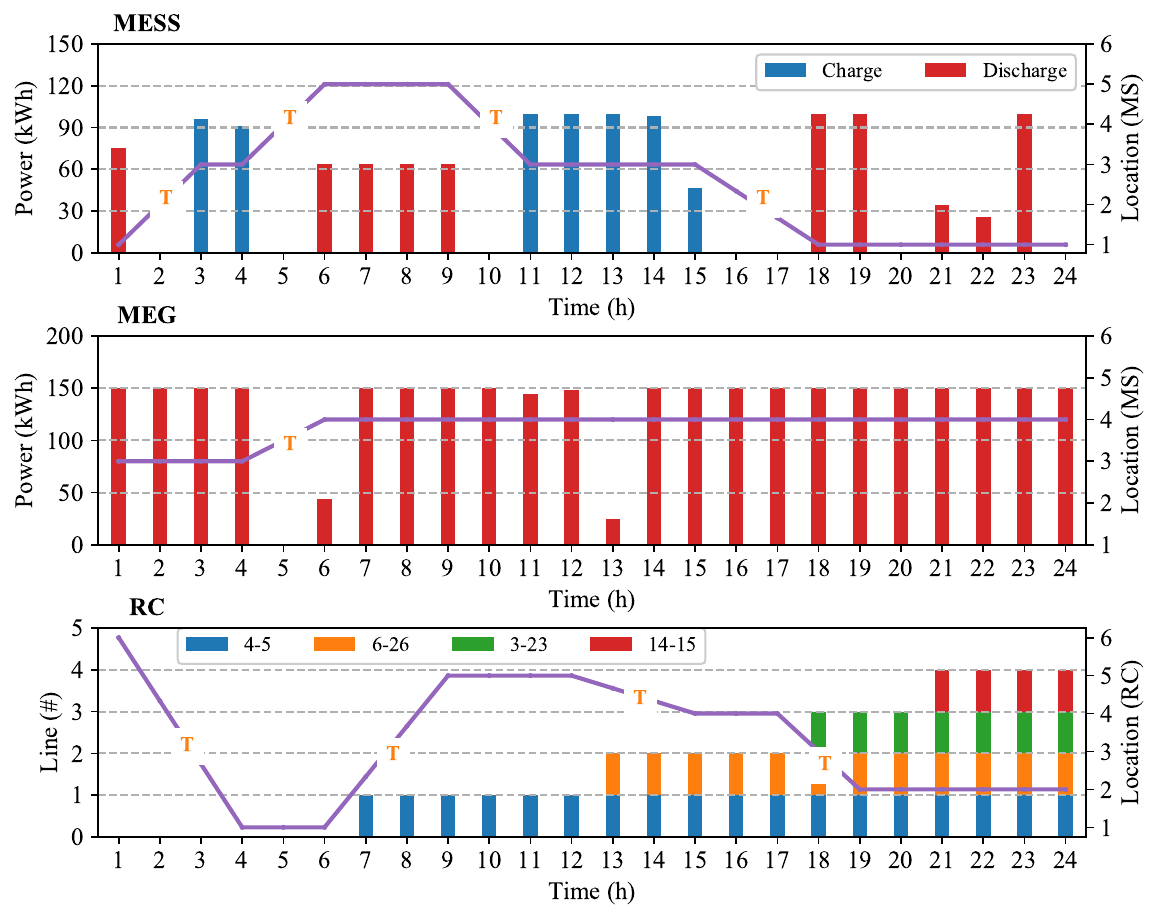}
\caption{Dispatch behaviors of MESS, MEG and RC.}
\label{fig:33bus rc mps}
\end{figure}

\begin{table}[h!]
\footnotesize
\centering
\setlength{\abovecaptionskip}{10pt}
\renewcommand\arraystretch{1.00}
\caption{Contribution of MESS, MEG and RC to Load Restoration in 33-Bus Power Network}
\setlength{\tabcolsep}{2.48mm}{
\begin{tabular}{ | l | c | c | c | }
      \hline
      Agent & MESS ($|P^{esd}|$) & MEG ($P^{eg}$) & RC ($|P^{rc}|$) \\
      \hline
      Quantity (kWh) & 692 & 3,211 & 6,363 \\
      \hline
\end{tabular}}
\label{table:33bus contribution}
\end{table}

\subsubsection{Dispatch of MPSs and RCs}
We first examine the dispatch behaviors of three MESS, MEG and RC agents, as depicted in Fig. \ref{fig:33bus rc mps}. As for MESS in Fig. \ref{fig:33bus rc mps}-(a), its routing behaviors are between MSs 1, 3 and 5. Specifically, the MESS chooses to discharge power at MSs 1 and 5 for demand supply, since both MS 1 at bus 2 and MS 5 at bus 25 connect with essential loads. Additionally, the discharging behaviors of MESS mainly occur at the periods of morning and night, when demand is relatively high. Now, let us look at the charging behaviors of MESS when it runs out of energy. The first charge occurs in the evening at MS 3 where MEG chooses to connect for power supply during the first few hours, as shown in Fig. \ref{fig:33bus rc mps}-(b). Such phenomena also exhibits the coordination effect of MESS and MEG in both mobility and flexibility. The second charge occurs in the mid-day when free PV resources are abundant. Furthermore, the interesting results can be found that it takes MESS 2 hours (15:01-17:00) to travel from MS 3 to MS 1 in the afternoon while taking only 1 hour (1:01-2:00) from MS 1 to MS 3 in the morning. This is because the serious road congestion happening in the afternoon leads to another hour traveling time. On the other hand, MEG chooses to connect with MS 3 at bus 32 and MS 4 at bus 14 for power supply, since MS 4 is connected with essential load and one serious damage happens around bus 14. As for RC in Fig. \ref{fig:33bus rc mps}-(c), it chooses to repair the damaged lines $4-5$, $6-26$, $3-23$ and $14-15$ sequentially. After these four lines are all repaired, RC has run out of its resources and is incapable of repairing more. It is also mentioned here the reason why RC firstly repairs line $4-5$ is that repairing this line can restore the associated power flow, in which bus 2 is connected with essential load. In this case, there is no need for MEG to connect with MS 1 at bus 2 towards resilience enhancement. After analyzing the dispatch behaviours of three MESS, MEG and RC agents, we also summarize the overall contribution of each agent to the system resilience in Table \ref{table:33bus contribution}. RC enhancing power flow contributes the most, followed by MEG and MESS. As such, it can be concluded that H2MAPPO successfully learns the reasonable dispatch behaviors for MESS, MEG and RC agents with the objective of providing resilience.

\begin{table}[h]
\footnotesize
\centering
\setlength{\abovecaptionskip}{10pt}
\renewcommand\arraystretch{1.00}
\caption{Switch Operations for Power Network Reconfiguration}
\begin{tabular}{|c|l|}
    \hline
    Time period (hr) & Switch Operations \\
    \hline
    1 & close 8-21, 9-15, 12-22, 18-33, 25-29 \\
    \hline
    13 & open 25-29 \\
    \hline
    18 & open 26-27, 9-15 and 12-22 \\
    \hline
    22 & open 8-21 \\
    \hline
\end{tabular} 
\label{tab:33bus switch}
\vspace{0.5em}
\end{table}

\subsubsection{Switch Operations}
Besides MESS, MEG and RC, switch operations that consider dynamic network reconfiguration can also help enhance resilience. It can be observed from Table \ref{tab:33bus switch} that all five tie lines are closed to restore the obstructed power flow at the beginning of the day. But after RC gradually repairs the damaged lines, some switches open up to ensure the power network radiality. As a result, the resilience can be further enhanced via the smart network reconfiguration. 

\begin{figure}[h!]
\centering
\includegraphics[width=0.485\textwidth]{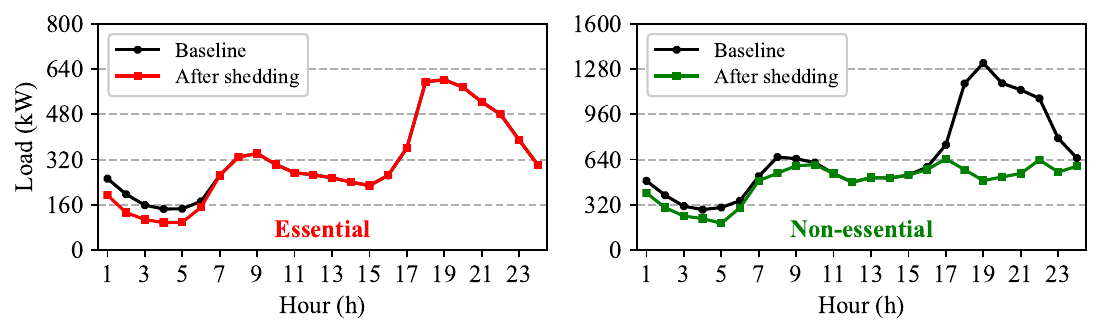}
\caption{Aggregated baseline and load after shedding in 33-bus system.}
\label{fig:33bus shedding} 
\end{figure}

\begin{table}[h!]
\footnotesize
\centering
\setlength{\abovecaptionskip}{10pt}
\renewcommand\arraystretch{1.00}
\caption{Load Shedding Quantity and Cost in 33-Bus Power Network}
\setlength{\tabcolsep}{2.98mm}{
\begin{tabular}{ | l | c | c | }
 	  \hline
      Performance & Essential load & Non-essential load \\
      \hline
      Quantity (kWh) & 291 & 4,217 \\
      \hline
      Cost (\pounds) & 728 & 6,326 \\
      \hline
\end{tabular}}
\label{table:33bus cost}
\end{table}

Finally, load conditions for both essential and non-essential types are compared in Fig.~\ref{fig:33bus shedding}. Overall, the resilience enhancement for essential loads exhibits better performance than that for non-essential loads, respectively causing 291 kWh and 4,217 kWh total load shedding quantity, as compared in Table \ref{table:33bus cost}. Thus, the system needs to pay serious cost for non-essential loads (6,326 \pounds) compared to the essential loads (728 \pounds).

\subsection{Test Results in Modified IEEE 69-Bus Power Network}
\label{sec:V.D}
\begin{figure}[h!]
\centering
\includegraphics[width=0.485\textwidth]{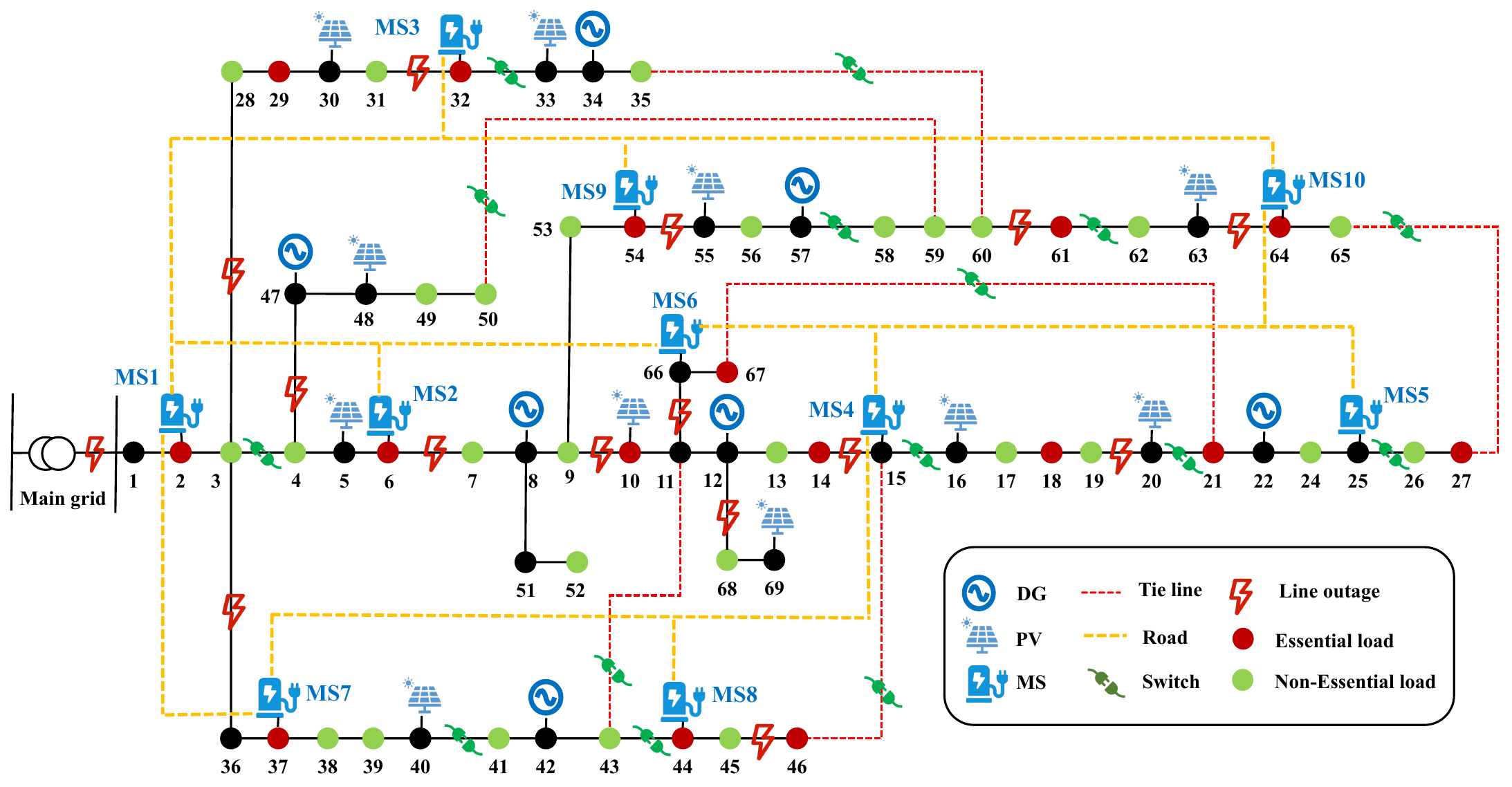}
\caption{The coupled 69-bus power-transport network.}
\label{fig:69bus}
\end{figure}

This section serves as a further demonstration of the proposed H2MAPPO on scalability. Thus, a modified IEEE 69-bus power network is introduced, which includes 7 DGs, 11 PVs and 10 MSs, while 4 MESSs, 4 MEGs and 4 RCs are employed for load restoration. The detailed structure of the coupled power-transport network can be found in Fig. \ref{fig:69bus}.

\begin{figure}[h!]
\centering
\includegraphics[width=0.485\textwidth]{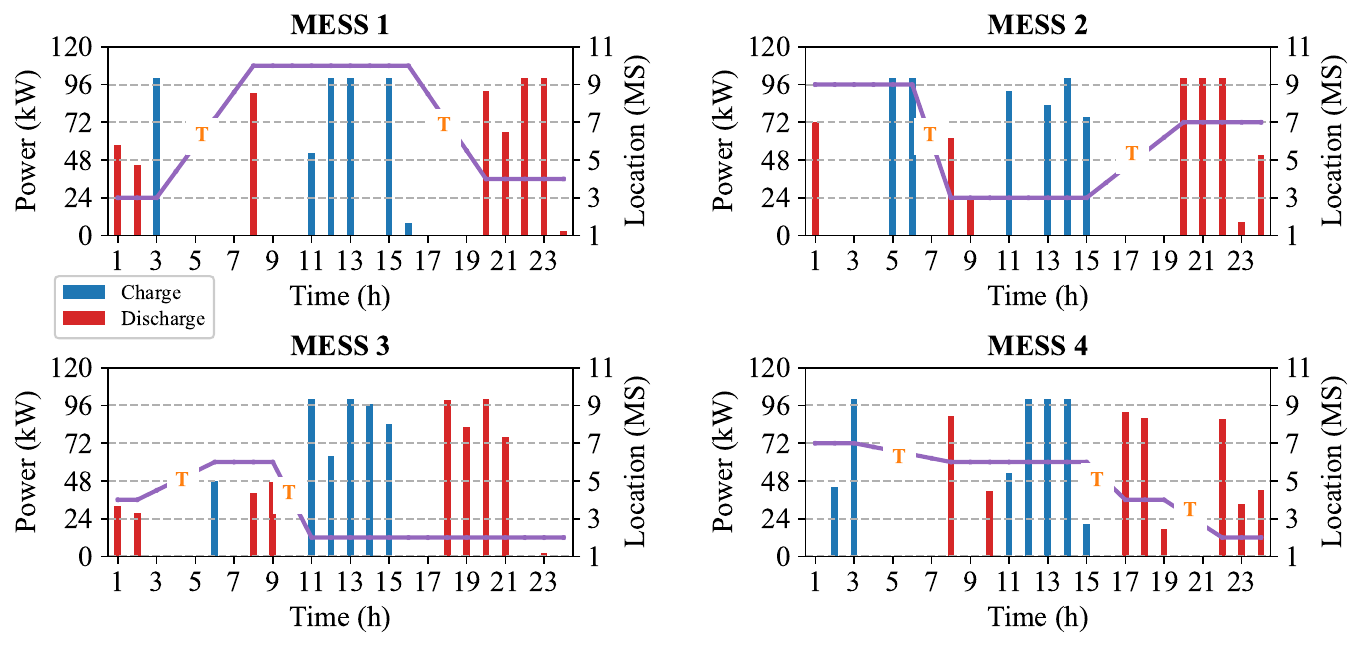}
\caption{Dispatch behaviors of MESSs in the modified 69-bus system.}
\label{fig:69bus mess}
\end{figure}

\begin{figure}[h!]
\centering
\includegraphics[width=0.485\textwidth]{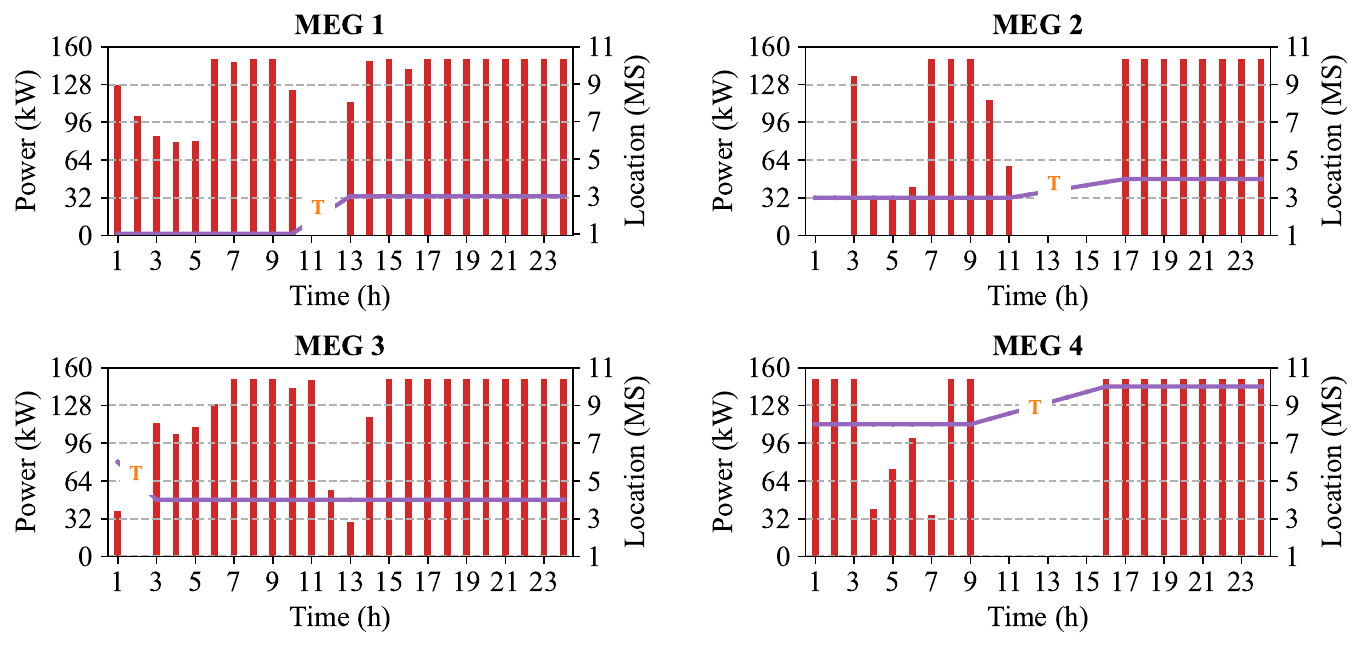}
\caption{Dispatch behaviors of MEGs in the modified 69-bus system.}
\label{fig:69bus meg}
\end{figure}

\begin{figure}[h!]
\centering
\includegraphics[width=0.485\textwidth]{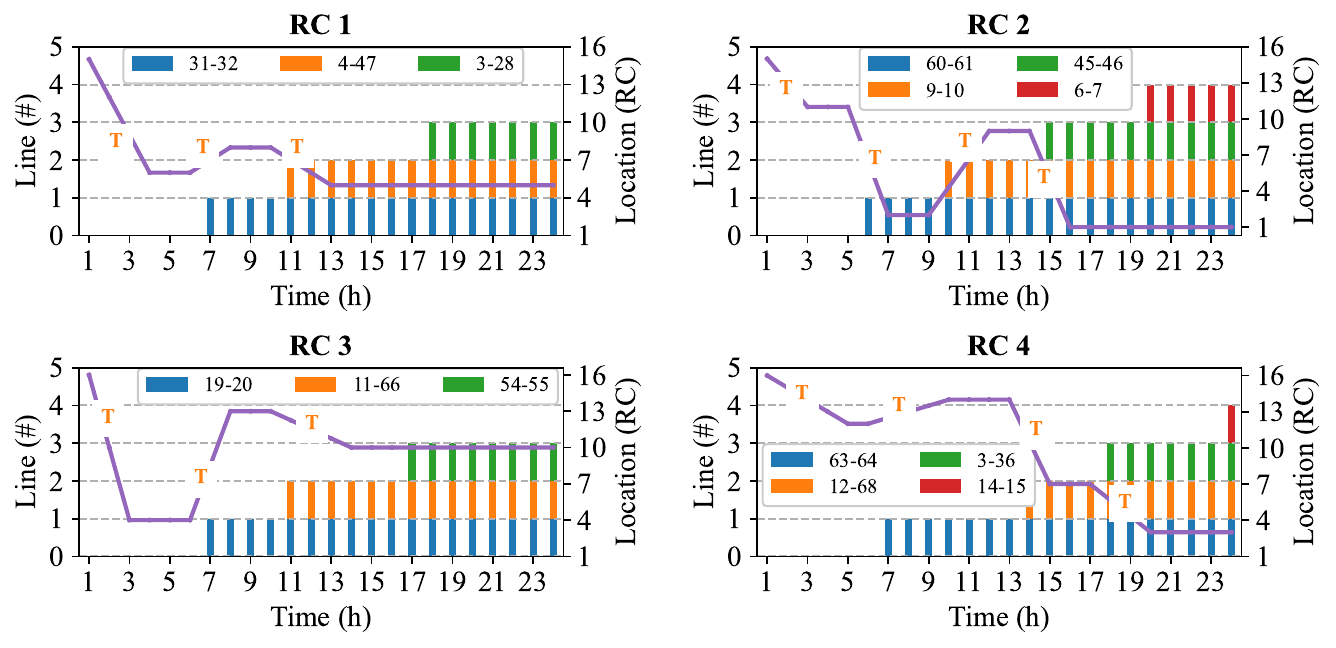}
\caption{Dispatch behaviors of RCs in the modified 69-bus system.}
\label{fig:69bus rc}
\end{figure}

\begin{figure}[h!]
\centering
\includegraphics[width=0.485\textwidth]{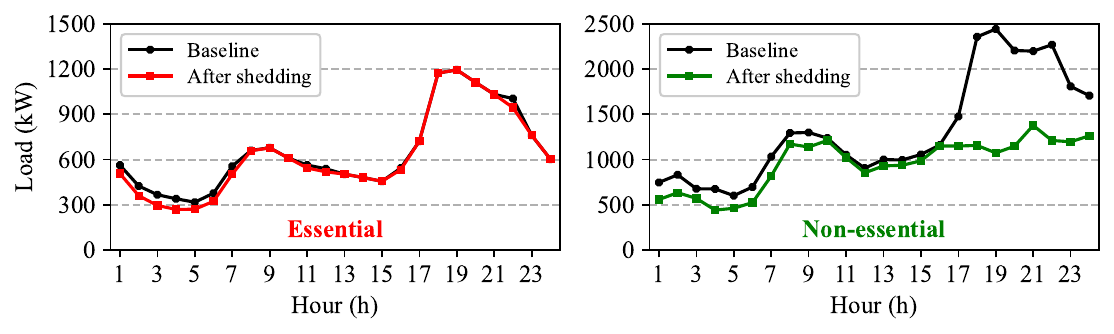}
\caption{Aggregated baseline and load after shedding in 69-bus system.}
\label{fig:69bus shedding} 
\end{figure}

\begin{table}[t!]
\footnotesize
\centering
\setlength{\abovecaptionskip}{10pt}
\renewcommand\arraystretch{1.00}
\caption{Load Shedding Quantity and Cost in 69-Bus Power Network}
\setlength{\tabcolsep}{2.98mm}{
\begin{tabular}{ | l | c | c | }
 	  \hline
      Performance & Essential load & Non-essential load \\
      \hline
      Quantity (kWh) & 527 & 8,726 \\
      \hline
      Cost (thous.\pounds) & 1,318 & 13,089 \\
      \hline
\end{tabular}}
\label{table:69bus cost}
\end{table}

\begin{table}[h!]
\footnotesize
\centering
\setlength{\abovecaptionskip}{10pt}
\renewcommand\arraystretch{1.00}
\caption{Contribution of MESSs, MEGs and RCs to Load Restoration in 69-Bus Power Network}
\setlength{\tabcolsep}{2.48mm}{
\begin{tabular}{ | l | c | c | c | }
 	  \hline
      Agent & MESS ($|P^{esd}|$) & MEG ($P^{eg}$) & RC ($|P^{rc}|$) \\
      \hline
      Quantity (kWh) & 2,068 & 10,299 & 26,495 \\
      \hline
\end{tabular}}
\label{table:69bus contribution}
\vspace{0.5em}
\end{table}

The dispatch behaviors of MESSs, MEGs, and RCs are illustrated in Figs. \ref{fig:69bus mess}, \ref{fig:69bus meg}, \ref{fig:69bus rc}, respectively. Similar to the results in the 33-bus power network, MESSs and MEGs are coordinating with each other to provide power supply for the modified 69-bus power network towards timely load restoration. On one hand, 4 MESSs choose to charge power in the midday at MSs 2, 3, 6, 10 respectively due to their nearby high PV penetration, while discharging power in the evening at MSs connected with essential loads, e.g., MESS 1 at MS 4, MESS 2 at MS 7, MESS 3 at MS 2, and MESS 4 at MSs 2 and 4. On the other hand, MEGs move to the MSs connected with essential loads for power supply, e.g., MEG 1 at MSs 1 and 3, MEG 2 at MSs 3 and 4, MEG 3 at MS 4, and MEG 4 at MSs 8 and 10. Specifically, it can be found that MEG 3 moves to MS 4 and stays there all day, while MEG 2, and MESSs 1 and 4 are also connected with MS 4 in the evening when the load level reaches its peak. It is because the line outage occurring on line $45-46$ causes the isolation of the essential load at bus 46, while MPSs with black-start capabilities can energize bus 46 and provide power supply for this essential load via MS 4 and the smart switch operation on line $15-46$. 

Furthermore, 4 RCs repair the damaged lines sequentially, where lines $31-32$, $60-61$, $19-20$, and $63-64$ are repaired in the first order. On one hand, repairing line $31-32$ can restore the power supply to the area around buses 28-31 including one essential load, since the only power source in this area is the grid-following PV at bus 30 without black-start capability. On the other hand, repairing lines $60-61$, $19-20$, and $63-64$ can restore the connections between the areas around buses 61-65 and 20-27 (including several essential loads) and the main power network. In particular, buses 61-63 can only be energized after line $60-61$ is repaired at 6:00, because of the lack of black-start capability in this area (grid-following PV at bus 63). 

From the perspective of their coordination effect, it can be found that RCs firstly focus on repairing damaged components on the right part of the power network, while MPSs mainly choose to connect with MSs (e.g., MSs 1, 3, 4, and 8) on the left part of the power network. For instance, MEG 1 with black-start capability is connected with MS 1 to energize the area around buses 1-6 and provide power supply in the first 10 hours and then move to MS 3 when the line $4-47$ is repaired by RC 1 at 11:00. In this case, the DG unit at bus 47 is capable of providing power supply and black-start capability. Such coordination behaviors ensure that most buses in the power network can be energized quickly and enable fast load restoration.

Similar to the 33-bus system, the performance on providing resilience for essential loads is better than that for non-essential loads, as compared in Fig. \ref{fig:69bus shedding} and Table \ref{table:69bus cost}. The cost for non-essential loads is thus much higher than that for essential loads. Furthermore, it can be observed in Table \ref{table:69bus contribution} that RCs enhancing power flow are also expected to contribute the most, followed by MEGs and MESSs. These results further validate the effectiveness of the proposed H2MAPPO in supporting system resilience for a large-scale system.

\section{Conclusions and Future Work}
\label{sec:VI}
This paper proposes a novel MARL method to address the coordinated dispatch problem of MPSs and RCs for MG load restoration. The proposed MARL method is characterized by a hierarchical architecture and a hybrid action domain including both discrete routing and continuous scheduling actions. The resilience-driven coordinated dispatch problem of MPSs and RCs is formulated as a Dec-POMDP, rendering a decentralized fashion and capturing the system dynamics of the coupled power-transport networks. Additionally, uncertainties related to renewables, demand, traffic volumes, and line outages are encompassed in the MARL training procedure. Experiment results based on two power networks (IEEE 33-bus and IEEE 69-bus) demonstrate the effectiveness of the coordinated dispatch of MPSs and RCs on restoring loads and enhancing resilience, while the outstanding performance of the proposed MARL method in optimality, stability, and scalability is testified, compared to the state-of-the-art MARL and optimization methods.

Future work aims at enhancing the studied problem from three directions. 
First, this paper focuses on the short-term daily load restoration problem. However, extreme events may have a long-term impact on the power system infrastructure. As a result, the first future extension is applying the proposed MARL method to a long-term load restoration problem that could last for several days/weeks. In this case, the episodic horizon will be expanded, e.g., 168 time steps for a 7-day episode with 1 hour per time step, while more efficient data sampling techniques could be applied to deal with the longer episodic horizon. Furthermore, these MPS and RC agents should be capable of learning multi-task policies within a single episode. For example, the re-filling process of resources or fuels that captures the influence of working time and resource availability can be considered in the RL training process by modifying the Dec-POMDP setup, e.g., appropriately adding a penalty term to the reward function to avoid the long-term working period.
Second, this paper focuses on the routing and scheduling/repairing behaviors of MPSs and RCs, while their pre-allocation problem is not considered. However, the effective pre-allocation of these mobile and flexible sources can further improve system resilience. As a result, the second future extension is developing an optimal pre-allocation scheme for their initial positions and numbers of mobile sources towards system resilience enhancement.
Third, this paper models the uncertainties of RL environment in terms of load profiles, PV generation, traffic congestion, and line outages, while the uncertainties of mobile sources themselves are not captured. However, in real-world applications, the mobile sources are also characterized by various uncertainties, e.g., the repair time of RCs, the charging losses of MESSs, and the fuel consumption of MEGs. As a result, the third future extension is taking the uncertainties of mobile sources into account and developing more realistic dispatch behaviors towards load restoration. 

\bibliographystyle{IEEEtran}
\bibliography{References.bib}
\ifCLASSOPTIONcaptionsoff
  \newpage
\fi

\begin{IEEEbiography}[{\includegraphics[width=1in,height=1.25in,clip,keepaspectratio]{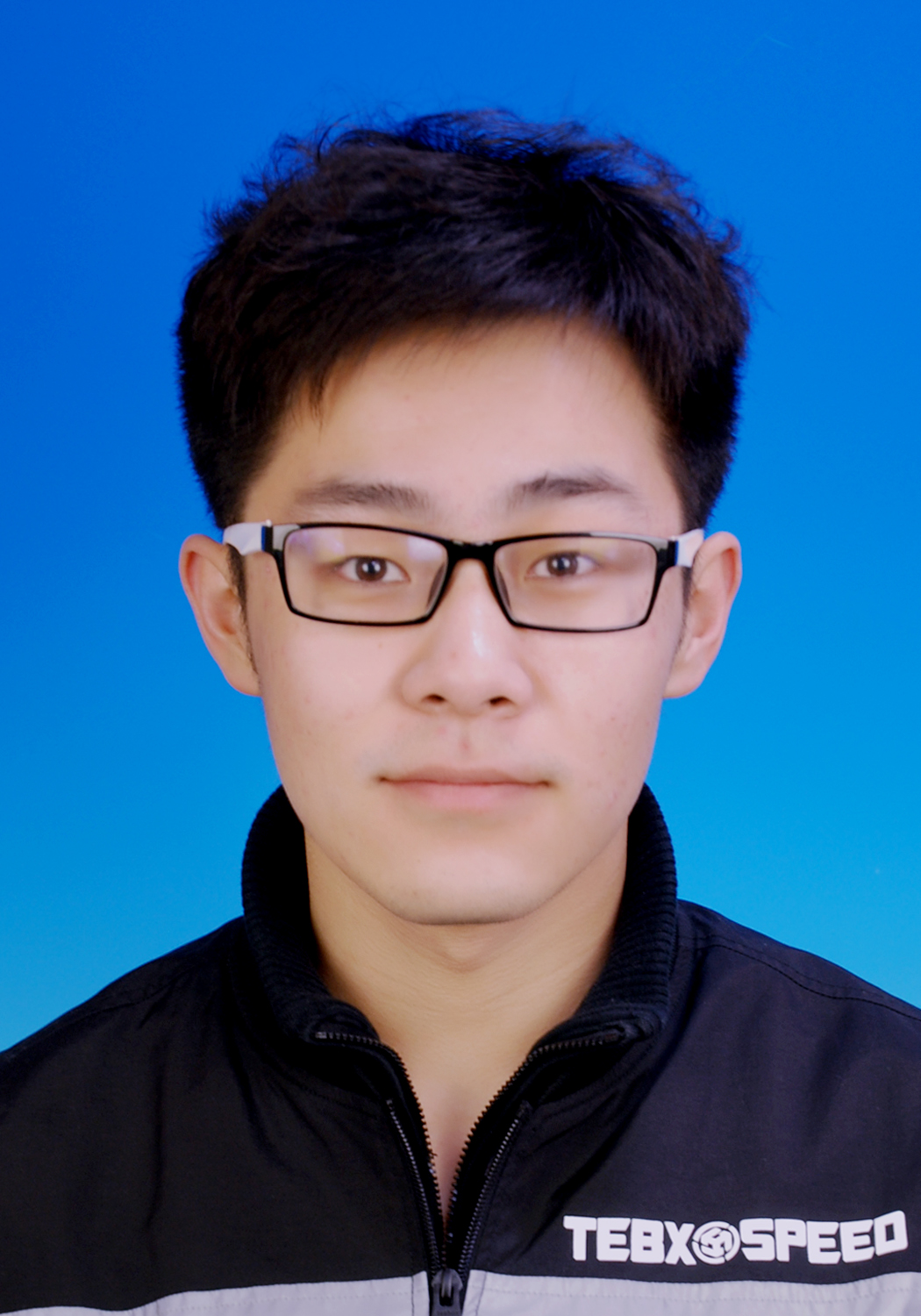}}]{Yi Wang}
received the Ph.D. degree from the Department of Electrical and Electronic Engineering at Imperial College London, U.K., in 2022. He is currently employed as a Research Associate in the Department of Electrical and Electronic Engineering at Imperial College London. His research interests include mathematical programming and learning approaches applied to the planning and operation of networked microgrids, the resilience enhancement of future power systems, and multi-energy system integration.
\end{IEEEbiography}

\begin{IEEEbiography}[{\includegraphics[width=1in,height=1.25in,clip,keepaspectratio]{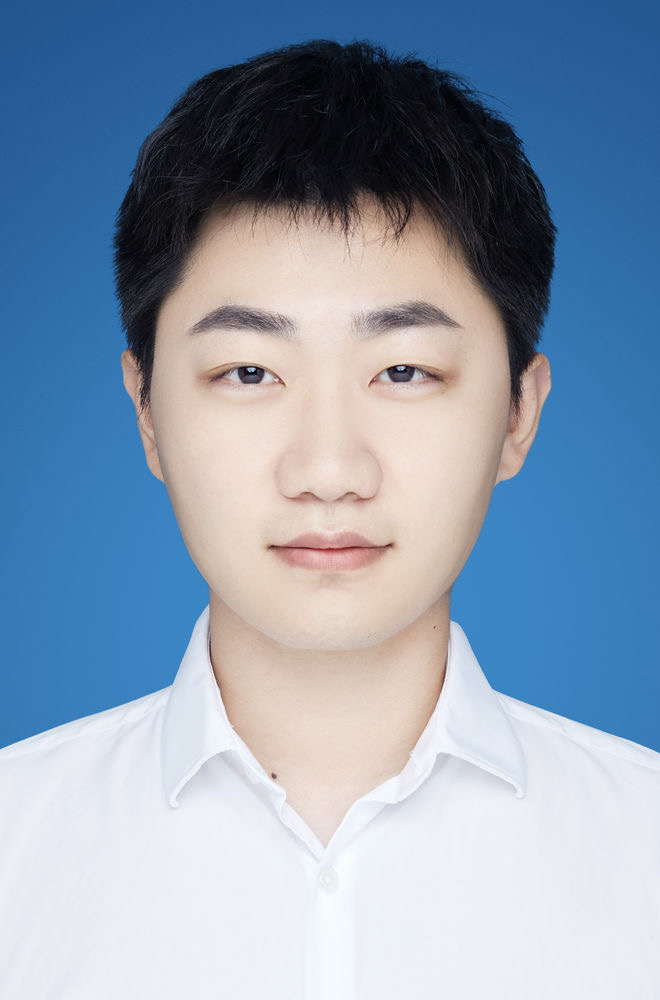}}]{Dawei Qiu}
received the B.Eng. degree in Electrical and Electronic Engineering from Northumbria University, U.K., in 2014, the M.Sc. degree in Power System Engineering from University College London, U.K., in 2015, and the Ph.D. degree in Electrical Engineering Research from Imperial College London, U.K., in 2020. He is currently employed as a Research Associate in the Department of Electrical and Electronic Engineering at Imperial College London. His research focuses on the development and application of decentralized and market-based approaches to electricity market, peer-to-peer energy trading, multi-energy system integration, microgrid resilience control, and vehicle-to-grid flexibility. In particular, he has a strong background in game theoretic modelling and reinforcement learning approaches.
\end{IEEEbiography}

\begin{IEEEbiography}[{\includegraphics[width=1in,height=1.25in,clip,keepaspectratio]{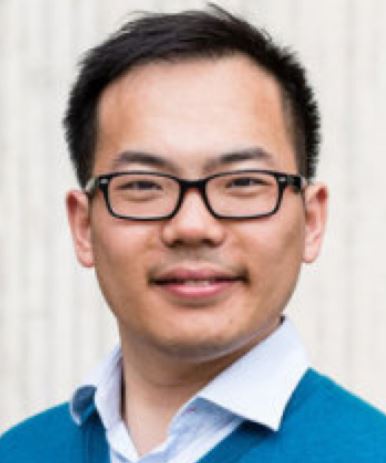}}]{Fei Teng}
received the B.Eng in Electrical Engineering from Beihang University, China, in 2009, and the M.Sc. and Ph.D. degrees in Electrical Engineering from Imperial College London, U.K., in 2010 and 2015. Currently, he is a Lecturer in the Department of Electrical and Electronic Engineering, Imperial College London, U.K. His research focuses on cyber-physical modeling, optimization and data analytics of power systems.
\end{IEEEbiography}

\begin{IEEEbiography}[{\includegraphics[width=1in,height=1.25in,clip,keepaspectratio]{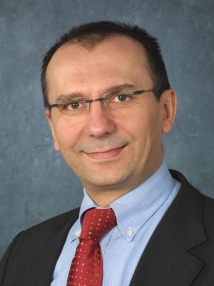}}]{Goran Strbac}
is a Professor of Energy Systems at Imperial College London, London, U.K. He led the development of novel advanced analysis approaches and methodologies that have been extensively used to inform industry, governments, and regulatory bodies about the role and value of emerging new technologies and systems in supporting cost effective evolution to smart low carbon future. He is currently the Director of the joint Imperial-Tsinghua Research Centre on Intelligent Power and Energy Systems, Leading Author in IPCC WG 3, Member of the European Technology and Innovation Platform for Smart Networks for the Energy Transition, and Member of the Joint EU Programme in Energy Systems Integration of the European Energy Research Alliance.
\end{IEEEbiography}

\end{document}